\documentstyle[preprint,floats,tighten,prd,aps]{revtex}
\input epsf.tex
\def\DESepsf(#1 width #2){\epsfxsize=#2 \epsfbox{#1}}
\begin{document}
\draft

\title{Techniques for QCD calculations by numerical integration}
\author{ Davison E.\ Soper}
\address{Institute of Theoretical Science, 
University of Oregon, Eugene, OR 97403}
\date{10 October 1999}
\maketitle

\begin{abstract}
Calculations of observables in quantum chromodynamics are typically
performed using a method that combines numerical integrations
over the momenta of final state particles with analytical integrations
over the momenta of virtual particles. I describe the most important
steps of a method for performing all of the integrations numerically.
\end{abstract}




\section{Introduction}
\label{sec:intro}

This paper concerns a method, which was introduced in
\cite{beowulfprl}, for performing perturbative calculations in
quantum chromodynamics (QCD) and other quantum field theories. The
method is intended for calculations of quantities in which one
measures something about the hadronic final state produced in a
collision and in which the observable is infrared safe -- that is,
insensitive to long-distance effects. Examples include jet cross
sections in hadron-hadron and lepton-hadron scattering and in $e^+ e^-
\to {\it  hadrons}$. There have been many calculations of this kind 
carried out at next-to-leading order in perturbation theory. These
calculations are based on a method introduced by Ellis, Ross, and
Terrano \cite{ERT} in the context of $e^+ e^- \to {\it  hadrons}$.
Stated in the simplest terms, the Ellis-Ross-Terrano method is to do
some integrations over momenta $\vec \ell_i$ analytically, others
numerically. In the method discussed here, one does all of these
integrations numerically. Evidently, if one performs all of the
integrations numerically, one gains flexibility to quite easily modify
the integrand. There may be other advantages, as well as some
disadvantages, to the numerical integration method compared to the
numerical/analytical method. 

In this paper, I address only the process $e^+ e^- \to {\it  hadrons}$.
I discuss three-jet-like infrared safe observables at next-to-leading
order, that is order $\alpha_s^2$. Examples of such observables
include the thrust distribution and the fraction of events that have
three jets.

The main techniques of the numerical integration method for $e^+ e^-
\to {\it  hadrons}$ were presented briefly in \cite{beowulfprl}. The
principle purpose of this paper is to explain in detail some of the
most important of these techniques. In the numerical/analytical
method, one has to work hard to implement the cancellation of
``collinear'' and ``soft'' divergences that occur in the integrations.
In the numerical method, as we will see, this cancellation happens
automatically. On the other hand, in the completely numerical method
one has the complication of having to deform some of the integration
contours into the complex plane. We will see how to do this
deformation. In both the numerical/analytical method and the
completely numerical method, one must arrange that the density of
integration points is singular near a soft gluon singularity of the
integrand (even after cancellations).  However, the precise behavior
of the densities needed in the two cases is different. We will see
what is needed in the numerical method.

These techniques are presented in
Secs.~\ref{sec:themodel}-\ref{sec:MonteCarlo}. They are illustrated in
Sec.~\ref{sec:example} with a numerical example. Although a full
understanding of the example requires the preceding sections, the
reader may want to look briefly at Sec.~\ref{sec:example} before starting
on Secs.~\ref{sec:themodel}-\ref{sec:MonteCarlo}. A brief summary of
techniques not presented in detail in this paper is given in
Sec.~\ref{sec:other}.

In \cite{beowulfprl}, I presented results from a concrete
implementation of the numerical method in computer code. Since then,
one logical error in the code has been discovered and fixed and the
performance of the program has been improved. Results from the
improved code \cite{beowulf} are presented in Sec.~\ref{sec:conclusions}. 

Let us begin with a precise statement of the problem. We consider an
observable such as a particular moment of the thrust distribution.
The observable can be expanded in powers of $\alpha_s/\pi$,
\begin{equation}
\sigma = \sum_n 
\sigma^{[n]},
\hskip 1 cm
\sigma^{[n]} \propto \left(\alpha_s / \pi\right)^n\,.
\end{equation}
The order $\alpha_s^2$ contribution has the form
\begin{eqnarray}
\sigma^{[2]} &=&
{1 \over 2!}
\int d\vec k_1 d\vec k_2\
{d \sigma^{[2]}_2 \over d\vec k_1 d\vec k_2}\
{\cal S}_2(\vec k_1,\vec k_2)
\nonumber\\
&&+
{1 \over 3!}
\int d\vec k_1 d\vec k_2 d\vec k_3\
{d \sigma^{[2]}_3 \over d\vec k_1 d\vec k_2 d\vec k_3}\
{\cal S}_3(\vec k_1,\vec k_2,\vec k_3)
\label{start}\\
&&
+
{1 \over 4!}
\int d\vec k_1 d\vec k_2 d\vec k_3 d\vec k_4\
{d \sigma^{[2]}_4 \over d\vec k_1 d\vec k_2 d\vec k_3 d\vec k_4}\
{\cal S}_4(\vec k_1,\vec k_2,\vec k_3,\vec k_4).
\nonumber
\end{eqnarray}
Here the $d\sigma^{[2]}_n$ are the  order $\alpha_s^2$ contributions
to the parton level cross section, calculated with zero quark masses.
Each contains momentum and energy conserving delta functions. The $d
\sigma^{[2]}_n$ include ultraviolet renormalization in the
$\overline{\rm MS}$ scheme. The functions $\cal S$ describe the
measurable quantity to be calculated. We wish to calculate a
``three-jet-like'' quantity.  That is, ${\cal S}_2 = 0$. The
normalization is such that ${\cal S}_n = 1$ for $n = 2,3,4$ would give
the order $\alpha_s^2$ perturbative contribution the the total cross
section.  There are, of course, infrared divergences associated with
Eq.~(\ref{start}). For now, we may simply suppose that an infrared
cutoff has been supplied.

The measurement, as specified by the functions ${\cal S}_n$, is to be
infrared safe, as described in Ref.~\cite{KS}: the ${\cal S}_n$ are
smooth functions of the parton momenta and
\begin{equation}
{\cal S}_{n+1}(\vec k_1,\dots,\lambda \vec k_n,(1-\lambda)\vec k_n)
= 
{\cal S}_{n}(\vec k_1,\dots, \vec k_n)
\end{equation}
for $0\le \lambda <1$. That is, collinear splittings and soft
particles do not affect the measurement.

It is convenient to calculate a quantity that is dimensionless. Let the
functions ${\cal S}_n$ be dimensionless and eliminate the remaining
dimensionality in the problem by dividing by $\sigma_0$, the total
$e^+ e^-$ cross section at the Born level. Let us also remove the
factor of $(\alpha_s / \pi)^2$. Thus, we calculate
\begin{equation}
{\cal I} = {\sigma^{[2]} \over \sigma_0\ (\alpha_s/\pi)^2}.
\label{calIdef}
\end{equation}

Our problem is thus to calculate  ${\cal I}$. Let us now see how to 
set up this problem in a convenient form. We note that ${\cal I}$ is a
function of the c.m.\ energy $\sqrt s$ and the $\overline{\rm MS}$
renormalization scale $\mu$. We will choose $\mu$ to be proportional
to $\sqrt s$: $\mu = A_{UV} \sqrt s$. Then ${\cal I}$ depends on $A$.
But, because it is dimensionless, it is independent of $\sqrt s$. This
allows us to write
\begin{equation}
{\cal I} = \int_0^\infty d \sqrt s\ h(\sqrt s)\ 
{\cal I}(A_{UV},\sqrt s),
\end{equation}
where $h$ is any function with
\begin{equation}
\int_0^\infty d \sqrt s\ h(\sqrt s) = 1.
\label{rtsintegral}
\end{equation}

\begin{figure}
\centerline{\DESepsf(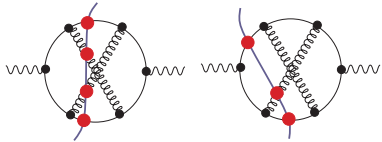 width 5 cm)}
\medskip
\caption{Two cuts of one of the Feynman diagrams that contribute to 
$e^+e^- \to {\it hadrons}$.}
\label{fig:cutdiagrams}
\end{figure}

The quantity $\cal I$ can be expressed in terms of cut Feynman
diagrams, as in Fig.~\ref{fig:cutdiagrams}. The dots where the parton
lines cross the cut represent the function ${\cal S}_n(\vec k_1,
\dots ,\vec k_n)$. Each diagram is a three loop diagram, so we have
integrations over loop momenta $\ell_1^\mu$, $\ell_2^\mu$ and
$\ell_3^\mu$. We first perform the energy integrations. For the graphs
in which four parton lines cross the cut, there are four mass-shell
delta functions $\delta (k_J^2)$. These delta functions eliminate the
three energy integrals over $\ell_1^0$, $\ell_2^0$, and $\ell_3^0$ as
well as the integral (\ref{rtsintegral}) over $\sqrt s$. For the
graphs in which three parton lines cross the cut, we can eliminate the
integration over $\sqrt s$ and two of the $\ell_J^0$ integrals. One
integral over the energy $E$ in the virtual loop remains. We perform
this integration by closing the integration contour in the lower half
$E$ plane. This gives a sum of terms obtained from the original
integrand by some algebraic substitutions, as we will see in the
following sections. Having performed the energy integrations, we are
left with an integral of the form
\begin{equation}
{\cal I} = \int d\vec \ell_1\,d\vec \ell_2\,d\vec \ell_3\
\sum_G\,\sum_C\,
g(G,C;\vec\ell_1,\vec\ell_2,\vec\ell_3).
\label{master}
\end{equation}
Here there is a sum over graphs $G$ (of which one is shown in
Fig.~\ref{fig:cutdiagrams}) and there is a sum over the possible cuts
of a given graph. 

The problem of calculating ${\cal I}$ is now set up in a convenient
form for calculation. If we were using the Ellis-Ross-Terrano method,
we would calculate some of the integrals in Eq.~(\ref{master})
numerically and some analytically. In the method described here, we
first perform certain contour deformations, then calculate all of the
integrals by Monte Carlo numerical integration. In the following
sections, we will learn the main techniques for performing the
integrations in Eq.~(\ref{master}). We will do this by studying a
simple model problem that will enable us to see the essential features
of the numerical method with as few extraneous difficulties as possible.

\section{A simplified model}
\label{sec:themodel}

\begin{figure}[htb]
\centerline{\DESepsf(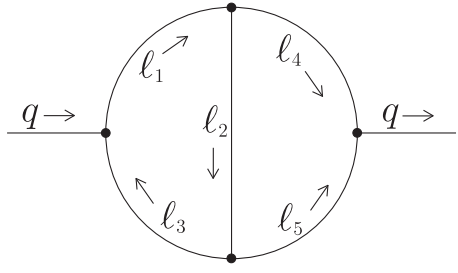 width 6 cm)}
\medskip
\caption{Diagram for a simple calculation.
All two and three parton cuts of this diagram in $\phi^3$ theory are
used, with a measurement function that gives the average transverse
energy in the final state.}
\label{fig:model}
\end{figure}

\begin{figure}[htb]
\centerline{\DESepsf(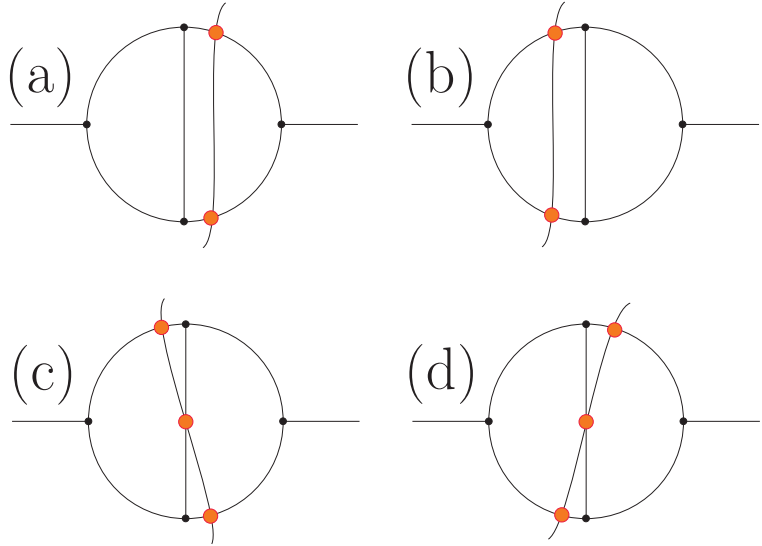 width 10 cm)}
\medskip
\caption{The two and three parton cuts of the simple $\phi^3$ diagram.}
\label{fig:cuts}
\end{figure}

In the following sections, we consider a simplified model in which all
complications that are not needed for a first understanding of the
numerical method are stripped away. The model is represented by the
graph shown in Fig.~\ref{fig:model}. There are contributions from all
of the two and three parton cuts of this diagram, as shown in
Fig.~\ref{fig:cuts}. Since QCD numerator functions do not play a
major role, we consider this graph in $\phi^3$ theory. Thus, also, we
can avoid the complications of ultraviolet renormalization. We
consider the incoming momentum $\vec q$ to be fixed and nonzero. We
calculate the integral of the graph over the incoming energy $q^0$.
This is analogous to the technical trick of integrating over $\sqrt s$
in the full three loop QCD calculation (see Sec.~\ref{sec:intro})
and serves to provide three energy integrations to perform against
three mass-shell delta functions for the three-parton cuts.

We need a nontrivial measurement function $\cal S$. As an example, we
choose to measure the transverse energy in the final state normalized
to the total energy:
\begin{eqnarray}
{\cal S}_2(\vec k_1,\vec k_2)&=&
(|\vec  k_{T,1}| + |\vec  k_{T,2}|)
/(|\vec  k_{1}| + |\vec  k_{2}|)
\nonumber\\
{\cal S}_3(\vec k_1,\vec k_2,\vec k_3)&=&
(|\vec  k_{T,1}| + |\vec  k_{T,2}| +  |\vec  k_{T,3}|)
/(|\vec  k_{1}| + |\vec  k_{2}| + |\vec  k_{3}|),
\end{eqnarray}
where $\vec  k_{T,j}$ is the part of the momentum $\vec  k_{j}$ of the
$j$th final state particle that is orthogonal to $\vec q$.

There are two loops in our diagram. We choose the independent loop
momenta to be $\ell_2^\mu$ and $\ell_4^\mu$. The other momenta are
understood to be expressed in terms of $\ell_2^\mu$, $\ell_4^\mu$, and
$q^\mu$.

Thus the example integral that we seek to calculate is
\begin{equation}
{\cal I} = {g^4 \over 2}
\int\! {d q^0 \over 2\pi}
\int\! {d^4\ell_2 \over (2\pi)^4}
\int\! {d^4\ell_4 \over (2\pi)^4}\ {\cal W}.
\label{samplestart}
\end{equation}
Here $g$ is the coupling, $1/2$ is the statistical factor for this
graph, and the integrand ${\cal W}$ consists of four parts, one for
each of the cuts in Fig.~\ref{fig:cuts}:
\begin{equation}
{\cal W} = {\cal W}_a +
{\cal W}_b + {\cal W}_c + {\cal W}_d,
\label{Wsum}
\end{equation}
where
\begin{eqnarray}
{\cal W}_a &=&
{i}
{\cal S}_2(\vec\ell_4,\vec\ell_5)\,
{1\over \ell_1^2 + i\epsilon}\,
{1\over \ell_2^2 + i\epsilon}\,
{1\over \ell_3^2 + i\epsilon}\,
(2\pi)\Delta(\ell_4)\,
(2\pi)\Delta(\ell_5),
\nonumber\\
{\cal W}_b &=&
{-i}
{\cal S}_2(\vec\ell_1,-\vec\ell_3)\,
(2\pi)\Delta(\ell_1)\,
{1\over \ell_2^2 - i\epsilon}\,
(2\pi)\Delta(-\ell_3)\,
{1\over \ell_4^2 - i\epsilon}\,
{1\over \ell_5^2 - i\epsilon},
\nonumber\\
{\cal W}_c &=&
{\cal S}_3(\vec\ell_1,-\vec\ell_2,\vec\ell_5)\,
(2\pi)\Delta(\ell_1)\,
(2\pi)\Delta(-\ell_2)\,
{1\over \ell_3^2}
{1\over \ell_4^2}\,
(2\pi)\Delta(\ell_5),
\nonumber\\
{\cal W}_d &=&
{\cal S}_3(\vec\ell_4,\vec\ell_2,-\vec\ell_3)\,
{1\over \ell_1^2}\,
(2\pi)\Delta(\ell_2)\,
(2\pi)\Delta(-\ell_3)\,
(2\pi)\Delta(\ell_4)\,
{1\over \ell_5^2}.
\label{Wparts}
\end{eqnarray}
Here we have used the notation 
\begin{equation}
\Delta(k) = \delta(k^2)\, \theta(k^0).
\end{equation}

\section{The integration over energies}
\label{sec:Eintegral}

\begin{figure}[htb]
\centerline{\DESepsf(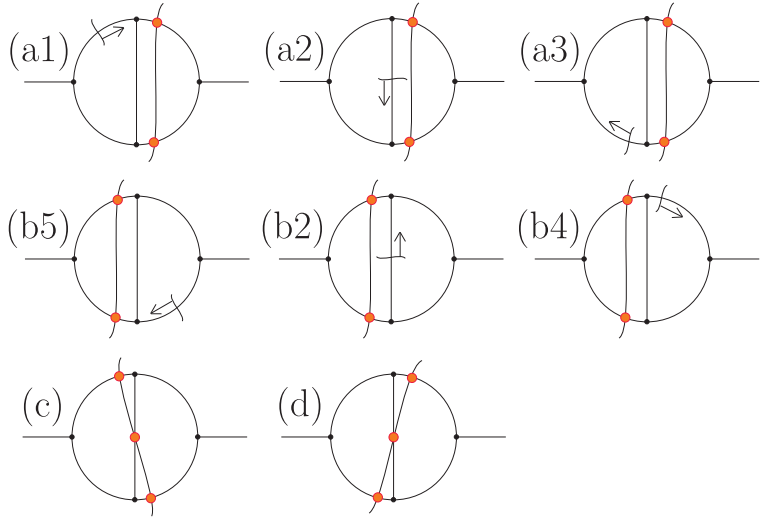 width 10cm)}
\medskip
\caption{The eight contributions to the sample diagram after
performing the energy integrations. The line through a propagator in
a loop indicates that this propagator is put on shell, with positive
energy flowing in the direction of the arrow. The direction for
positive energy flow around the loop depends on whether the contour
over loop energy is closed in the upper or the lower half plane.}
\label{fig:cuts8}
\end{figure}

We begin by performing the integrals over the energies in
Eq.~(\ref{samplestart}). In the case of three partons in the final
state, the three delta functions eliminate the three integrations. In
the case of two partons in the final state, the two delta functions
eliminate two of the energy integrations. This leaves one integration
over the energy that circulates around the virtual loop. There are
three poles in the upper half plane and three in the lower half plane.
Closing the contour in one half plane or the other gives three
contributions. Each of these contributions corresponds to putting one
of the particles in the loop on shell. Thus altogether there are
eight contributions to ${\cal I}$, as indicated in
Fig.~\ref{fig:cuts8}. We write ${\cal I}$ as
\begin{equation}
{\cal I} = {g^4 \over 2 (2\pi)^6}
\int\!d\vec \ell_4\
\int\!d\vec \ell_2\
{\cal G},
\label{theintegral}
\end{equation}
where the integrand ${\cal G}$ has eight parts:
\begin{equation}
{\cal G} = 
{\cal G}_{a1} + {\cal G}_{a2} + {\cal G}_{a3} +
{\cal G}_{b5} + {\cal G}_{b2} + {\cal G}_{b4} +
{\cal G}_c + {\cal G}_d.
\label{Gsum}
\end{equation}
The contributions to ${\cal G}$ are
\begin{eqnarray}
{\cal G}_{a1} &=&
{\cal S}_2(\vec\ell_4,\vec\ell_5)\,
{1\over 2|\vec\ell_1|}\,
{1\over (|\vec\ell_1| - |\vec\ell_4|)^2 -\vec\ell_2^{\,2} +
i\epsilon}\, {1\over (|\vec\ell_1| - |\vec\ell_4| - |\vec\ell_5|)^2 
  -\vec\ell_3^{\,2} + i\epsilon}\, 
{1\over 2|\vec\ell_4|}\,
{1\over 2|\vec\ell_5|},
\nonumber\\
{\cal G}_{a2} &=&
{\cal S}_2(\vec\ell_4,\vec\ell_5)\,
{1\over (|\vec\ell_2| + |\vec\ell_4|)^2 -\vec\ell_1^{\,2} +
i\epsilon}\, {1\over 2|\vec\ell_2|}\,
{1\over (|\vec\ell_2| - |\vec\ell_5|)^2 
  -\vec\ell_3^{\,2} + i\epsilon}\, 
{1\over 2|\vec\ell_4|}\,
{1\over 2|\vec\ell_5|},
\nonumber\\
{\cal G}_{a3} &=&
{\cal S}_2(\vec\ell_4,\vec\ell_5)\,
{1\over (|\vec\ell_3| + |\vec\ell_4| + |\vec\ell_5|)^2 
       - \vec\ell_1^{\,2} + i\epsilon}\, 
{1\over (|\vec\ell_3| + |\vec\ell_5|)^2 
       - \vec\ell_2^{\,2} + i\epsilon}\, 
{1\over 2|\vec\ell_3|}\,
{1\over 2|\vec\ell_4|}\,
{1\over 2|\vec\ell_5|},
\nonumber\\
{\cal G}_{b5} &=&
{\cal S}_2(\vec\ell_1,-\vec\ell_3)\,
{1\over 2|\vec\ell_1|}\,
{1\over (|\vec\ell_3| + |\vec\ell_5|)^2 
       - \vec\ell_2^{\,2} - i\epsilon}\, 
{1\over 2|\vec\ell_3|}\,
{1\over (|\vec\ell_1| + |\vec\ell_3| + |\vec\ell_5|)^2 
       - \vec\ell_4^{\,2} - i\epsilon}\, 
{1\over 2|\vec\ell_5|},
\nonumber\\
{\cal G}_{b2} &=&
{\cal S}_2(\vec\ell_1,-\vec\ell_3)\,
{1\over 2|\vec\ell_1|}\,
{1\over 2|\vec\ell_2|}\,
{1\over 2|\vec\ell_3|}\,
{1\over (|\vec\ell_1| + |\vec\ell_2|)^2 
       - \vec\ell_4^{\,2} - i\epsilon}\, 
{1\over (|\vec\ell_3| - |\vec\ell_2|)^2 
       - \vec\ell_5^{\,2} - i\epsilon} ,
\nonumber\\
{\cal G}_{b4} &=&
{\cal S}_2(\vec\ell_1,-\vec\ell_3)\,
{1\over 2|\vec\ell_1|}\,
{1\over (|\vec\ell_1| - |\vec\ell_4|)^2 
       - \vec\ell_2^{\,2} - i\epsilon}\,
{1\over 2|\vec\ell_3|}\,
{1\over 2|\vec\ell_4|}\,
{1\over (|\vec\ell_1| + |\vec\ell_3| - |\vec\ell_4|)^2 
       - \vec\ell_5^{\,2} - i\epsilon} ,
\nonumber\\
{\cal G}_c &=&
{\cal S}_3(\vec\ell_1,-\vec\ell_2,\vec\ell_5)\,
{1\over 2|\vec\ell_1|}\,
{1\over 2|\vec\ell_2|}\,
{1\over (|\vec\ell_2| + |\vec\ell_5|)^2 
       - \vec\ell_3^{\,2}}\,
{1\over (|\vec\ell_1| + |\vec\ell_2|)^2 
       - \vec\ell_4^{\,2}}\,
{1\over 2|\vec\ell_5|},
\nonumber\\
{\cal G}_d &=&
{\cal S}_3(\vec\ell_4,\vec\ell_2,-\vec\ell_3)\,
{1\over (|\vec\ell_2| + |\vec\ell_4|)^2 
       - \vec\ell_1^{\,2}}\,
{1\over 2|\vec\ell_2|}\,
{1\over 2|\vec\ell_3|}\,
{1\over 2|\vec\ell_4|}\,
{1\over (|\vec\ell_2| + |\vec\ell_3|)^2 
       - \vec\ell_5^{\,2}}.
\label{Gparts}
\end{eqnarray}

So far, the operations that we have performed have been purely
algebraic. They are evidently of a sort that can be easily implemented
in a computer program in an automatic fashion. We are left with an
integral over the loop momenta $\vec\ell_2$ and $\vec\ell_4$. We seek
to perform this integration numerically. However, the integrand $\cal
G$ has singularities, so it is not completely self-evident how to
proceed. It is to this question that we now turn.

\section{Cancellation of singularities}
\label{sec:cancellation}

In this section, we discuss the cancellation of singularities in a
numerical calculation of the integral in Eq.~(\ref{theintegral}).

\begin{figure}[htb]
\centerline{\DESepsf(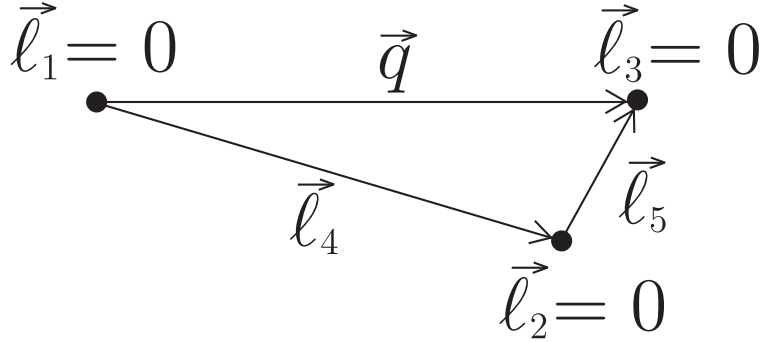 width 10 cm)}
\medskip
\caption{Space of loop momentum $\vec \ell_2$ for the virtual loop in
the graph of Fig.~\ref{fig:cuts}(a) for a representative choice of $\vec
q$, $\vec \ell_4$, and $\vec\ell_5 = \vec q - \vec\ell_4$.}
\label{fig:loopspace}
\end{figure}

Let us concentrate to begin with on the cut shown in
Fig.~\ref{fig:cuts}(a). Then there is a virtual loop consisting of the
propagators with momentum labels $\ell_1$, $\ell_2$ and
$\ell_3$. Recall that we are taking $\vec\ell_2$, and $\vec\ell_4$
as the independent loop momenta. Put the integration
over $\vec\ell_2$ inside the integration over $\vec\ell_4$. Then we can
consider $\vec\ell_4$ as fixed while $\vec\ell_2$ varies. 
Fig.~\ref{fig:loopspace} illustrates the space of the loop momentum
$\vec\ell_2$ for a particular choice of $\vec q$ and at a particular
point in the integration over $\vec\ell_4$. The origin of coordinates
is at the point labeled $\vec\ell_2 = 0$. The vector $\vec\ell_4$ is
indicated as an arrow with its head at $\vec\ell_2 = 0$. Then the point
$\vec\ell_1 = 0$ is at the tail of this vector, as indicated.  The
vector $\vec\ell_5 = \vec q - \vec\ell_4$ is indicated as an arrow with
its tail at $\vec\ell_2 = 0$. Then the point $\vec\ell_3 = 0$ is at the
head of this vector, as indicated. Finally, the vector $\vec q$ is
indicated as an arrow with its tail at $\vec\ell_1 = 0$.

Where are the singularities of the integrand for our graph? 
 
There is, first of all, a singularity when the momentum of any
propagator vanishes since there is always a contribution in which that
propagator is put on-shell, with a singularity $1/(2|\vec\ell|)$. 
Since an integration $\int d\vec \ell /(2|\vec\ell|)$ is convergent in
the infrared by two powers, these singularities do not cause much
difficulty. We simply have to choose a density of points with a
matching $1/|\vec\ell|$ singularity, as described later in 
Sec.~\ref{sec:MonteCarlo}. We do not discuss these singularities
further in this section.

The singularities of concern to us here are 

\smallskip\noindent
1) A collinear singularity at $\vec\ell_2 = - x \vec\ell_4$
 with $0 < x < 1$.

\smallskip\noindent
2) A collinear singularity at $\vec\ell_2 = x \vec\ell_5$
with $0 < x < 1$.

\smallskip\noindent
3) A soft singularity at $\vec\ell_2 = 0$.

\smallskip\noindent
4) A scattering singularity at $|\vec\ell_1| + |\vec\ell_3| =
|\vec\ell_4| + |\vec\ell_5|$.

\begin{figure}[htb]
\centerline{\DESepsf(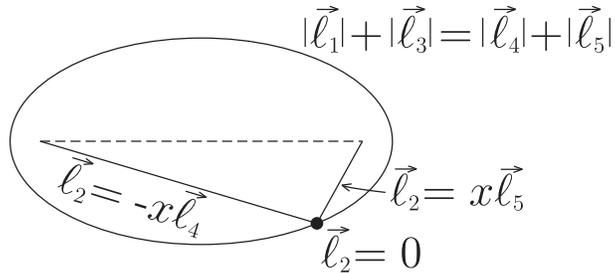 width 8 cm)}
\medskip
\caption{Locations of singularities of ${\cal G}_a$.}
\label{fig:singularities}
\end{figure}

\smallskip\noindent
The locations of these singularities are indicated in
Fig.~\ref{fig:singularities}.

\subsection{The collinear singularities}

In this subsection, we examine the collinear singularity at 
$\vec \ell_2 = - x \vec \ell_4$ with $0<x<1$. The principles that we
discover for this case will hold for the other collinear singularities
as well.

The terms ${\cal G}_{a1}$ and ${\cal G}_{c}$ in the integrand ${\cal
G}$, Eq.~(\ref{Gsum}), are singular along  the line $\vec \ell_2 = - x
\vec \ell_4$, $0<x<1$. In order to examine this singularity, let us
write ${\cal G}_{a1}$ and ${\cal G}_{c}$ as given in Eq.~(\ref{Gparts})
in the form
\begin{equation}
{\cal G}_{a1} = 
{ 1 \over 2 |\vec \ell_2 + \vec \ell_4|}\
{ 1 \over \left(E_2^{(a1)}\right)^2 - \vec \ell_2^{\,2}}\
{ 1 \over 2 |\vec \ell_4|}\
{\cal R}(E_1,E_2^{(a1)},E_5,\vec \ell_2,\vec \ell_4)\
{\cal S}_2(\vec\ell_4,\vec q - \vec\ell_4),
\end{equation}
\begin{equation}
{\cal G}_{c} = 
{ 1 \over 2 |\vec \ell_2 + \vec \ell_4|}\
{ 1 \over 2 |\vec \ell_2|}\
{ 1 \over \left(E_1 - E_2^{(c)}\right)^2 - \vec \ell_4^{\,2}}\
{\cal R}(E_1,E_2^{(c)},E_5,\vec \ell_2,\vec \ell_4)\
{\cal S}_3(\vec\ell_1,-\vec\ell_2,\vec q - \vec\ell_4).
\end{equation}
Here the first factors exhibit the denominators for the three
propagators that carry collinear momenta at the singularity,
${\cal R}$ denotes the rest of the Feynman graph, and the ${\cal S}$
functions are the measurement functions for the final state particles.
The functions ${\cal R}$ depend on the loop momenta $\vec\ell_2$
and  $\vec\ell_4$ and on three loop energies, which we take to be
$E_1 = \ell_1^0$, $E_2 = \ell_2^0$ and $E_5  = \ell_5^0$. The energies
are determined by the on-shell delta functions for the two
contributions. For $E_1$ and $E_5$, the values are the same for the two
contributions:
\begin{eqnarray}
E_1 &=& |\vec\ell_2 + \vec \ell_4|,
\nonumber\\
E_5 &=& |\vec q - \vec\ell_4|.
\end{eqnarray}
For $E_2$, the values are different:
\begin{eqnarray}
E_2^{(a1)}&=&|\vec\ell_2 + \vec \ell_4| - |\vec \ell_4|,
\nonumber\\
E_2^{(c)}&=& -|\vec\ell_2|.
\end{eqnarray}
In order to examine the behavior of ${\cal G}_{a1}$ and ${\cal G}_{c}$
near the singularity, let
\begin{equation}
\vec \ell_2 = - x \vec \ell_4 + \vec \ell_T ,
\end{equation}
where $\vec\ell_T \cdot \vec\ell_4 =0$. The singularity is at
$\vec\ell_T \to 0$. 

In ${\cal G}_{a1}$ the denominator $\left(E_2^{(a1)}\right)^2 - \vec
\ell_2^{\,2}$ vanishes at $\vec\ell_T \to 0$:
\begin{equation}
\left(E_2^{(a1)}\right)^2 - \vec \ell_2^{\,2} =
- { \vec\ell_T^{\,2} \over 1-x} 
\left( 1 + {\cal O}(\vec\ell_T^{\,2})\right).
\end{equation}
Thus there is a $1/\vec\ell_T^{\,2}$ singularity which would give a
logarithmically divergent result for the integral of ${\cal G}_{a1}$
alone. Altogether, the denominator factors for ${\cal G}_{a1}$ are
\begin{equation}
{ 1 \over 2 |\vec \ell_2 + \vec \ell_4|}\
{ 1 \over \left(E_2^{(a1)}\right)^2 - \vec \ell_2^{\,2}}\
{ 1 \over 2 |\vec \ell_4|}\
=
-{ 1 \over 4 \vec\ell_4^{\,2}}\
{ 1 \over \vec\ell_T^{\,2}}\
\left( 1 + {\cal O}(\vec\ell_T^{\,2})\right).
\label{denoma1}
\end{equation}

Let us now look at the denominator factors for ${\cal G}_{c}$. The
denominator $\left(E_1 - E_2^{(c)}\right)^2 - \vec \ell_4^{\,2}$
takes the form
\begin{equation}
\left(\left(E_1 - E_2^{(c)}\right)^2 - \vec \ell_4^{\,2}\right)^2 =
{ \vec\ell_T^{\,2} \over x(1-x)} 
\left( 1 + {\cal O}(\vec\ell_T^{\,2})\right),
\end{equation}
so that the denominator factors together take the form
\begin{equation}
{ 1 \over 2 |\vec \ell_2 + \vec \ell_4|}\
{ 1 \over 2 |\vec \ell_2|}\
{ 1 \over \left(E_1 - E_2^{(c)}\right)^2 - \vec \ell_4^{\,2}}\
=
{ 1 \over 4 \vec\ell_4^{\,2}}\
{ 1 \over \vec\ell_T^{\,2}}\
\left( 1 + {\cal O}(\vec\ell_T^{\,2})\right).
\label{denomc}
\end{equation}
Again, we have a $1/\vec\ell_T^{\,2}$ singularity.

Note, however, that the denominator factors in Eqs.~(\ref{denoma1})
and (\ref{denomc}) are equal except for their sign, up to corrections
that are not singular as $\vec\ell_T^{\,2} \to 0$. Thus if the
remaining factors ${\cal R}$ and ${\cal S}$ were exactly the same for
${\cal G}_{a1}$ and ${\cal G}_{c}$ there would be no singularity in
their sum.

We thus need to explore the matching of ${\cal R}$ and ${\cal S}$. The
two versions of ${\cal R}$ are the same functions with the same
arguments except for the fact that $E_2^{(c)} \ne E_2^{(a1)}$. However,
\begin{equation}
E_2^{(c)} = E_2^{(a1)} + {\cal O}(\vec\ell_T^{\,2}).
\end{equation}
Thus
\begin{equation}
{\cal R}(E_1,E_2^{(c)},E_5,\vec \ell_2,\vec \ell_4) =
{\cal R}(E_1,E_2^{(a1)},E_5,\vec \ell_2,\vec \ell_4) 
+ {\cal O}(\vec\ell_T^{\,2}).
\end{equation}
For the functions $\cal S$ used in our example, we have
\begin{equation}
{\cal S}_3(
(1-x)\vec\ell_4 +\vec\ell_T,
x\vec\ell_4 -\vec\ell_T,
\vec q - \vec\ell_4)
={\cal S}_2(\vec\ell_4, \vec q - \vec\ell_4)
+{\cal O}(\vec\ell_T^{\,2}).
\end{equation}
Using these matching equations we find that
\begin{equation}
{\cal G}_{a1} + {\cal G}_{c} = {\cal O}(1)
\end{equation}
as $\vec \ell_T \to 0$. There is no collinear singularity in $\cal G$.

How general is this result? First of all note that, in the part of the
argument not involving the measurement functions ${\cal S}$, we used
only the explicit structure of the denominators for three propagators
that meet at a vertex. In the limit in which the momenta carried on
these propagators become collinear, there is a cancellation of the
collinear singularity arising from these denominators. The three
propagators can be part of a much larger graph, and there can be
non-trivial numerator factors, as in QCD. All of the other factors
can be lumped into a function ${\cal R}$ and treated as above. Thus
this cancellation works in QCD as well as $\phi^3$ theory 
and it works for cut graphs with at most one virtual loop at any order
of perturbation theory. 

As for the measurement functions, in general we need to consider the
difference between the measurement functions with $n$ and $n+1$
particles in the final state,
\begin{equation}
F(\vec\ell_T) = 
{\cal S}_{n+1}(\vec k_1,\cdots,\vec k_{n-1},
x\vec k_n - \vec\ell_T,
(1-x)\vec k_n + \vec\ell_T)
-
{\cal S}_{n}(\vec k_1,\cdots,\vec k_{n-1},
\vec k_n).
\end{equation}
Assuming that $F$ is an analytic function of $\vec\ell_T$, it will have
an expansion around $\vec\ell_T = 0$ of the form
\begin{equation}
F(\vec\ell_T) =
a
+ b_i \cdot \ell_T^i
+ c_{ij}\ell_T^i \ell_T^j
+\cdots.
\end{equation}
Infrared safety requires that $a = 0$. If $b_i \ne 0$ then $F$
vanishes on a surface that intersects the point $\vec\ell_T = 0$.
Measurement functions ${\cal S}$ with this property would define an
infrared safe measurement, but I do not know of any
example in common use. More typically, $F$ is non-zero in a
neighborhood of $\vec \ell_T = 0$ while vanishing at $\vec \ell_T =
0$. Then both $a$ and the $b_i$ must vanish and the $c_{ij}$
should be a positive definite (or negative definite) matrix. Thus, for
typical measurement functions,
\begin{equation}
F(\vec\ell_T) = {\cal O}(\vec \ell_T^{\,2})
\end{equation}
as $\vec \ell_T \to 0$. Then the integrand does not have collinear
singularities.

For an atypical measurement function with $b_i \ne 0$, one would be
left with an integrable singularity of the form $\vec b \cdot
\vec\ell_T/\vec\ell_T^{\,2}$. The current version of the computer code
\cite{beowulf} has a mechanism to deal with this contingency, but I do
not discuss it here since I know of no case in which it is needed.

\subsection{The soft singularities}

In this subsection, we examine the soft singularity at $\vec \ell_2 =
0$. 

Let us concentrate to begin with on the cut graph shown in
Fig.~\ref{fig:cuts}(a). When we perform the integration over the energy
circulating in the virtual loop, there is a contribution from the term
in which the propagator carrying momentum $\ell_1^\mu$ is put on
shell, as in Fig.~\ref{fig:cuts8}(a1). This contribution is ${\cal
G}_{a1}$ in Eq.~(\ref{Gparts}). Let us examine this contribution
in the limit $\vec \ell_2 \to 0$. Expanding in powers of $\vec\ell_2$,
we have
\begin{equation}
\ell_1^0 = 
|\vec\ell_1| =
|\vec \ell_4 + \vec \ell_2| =
|\vec\ell_4| + |\vec\ell_2|\, \vec u_2\cdot \vec u_4 + \cdots,
\end{equation}
where we adopt the notation
\begin{equation}
\vec u_J = \vec\ell_J / |\vec\ell_J|.
\end{equation}
Then
\begin{equation}
\ell_2^2 = 
- |\vec\ell_2|^2\ [1 - (\vec u_2\cdot \vec u_4)^2]
+\cdots
\end{equation}
and
\begin{equation}
\ell_3^2 = 
2|\vec\ell_5|\,|\vec\ell_2|\
\vec u_2\cdot(\vec u_5 - \vec u_4)
+\cdots.
\end{equation}
Thus
\begin{eqnarray}
{\cal G}_{a1} &\sim& {\cal S}_2\,
{ 1 \over 2|\vec\ell_4|}\,
{ -1 \over |\vec\ell_2|^2\ [1 - (\vec u_2\cdot \vec u_4)^2]}\,
{ 1 \over 2|\vec\ell_5|\,|\vec\ell_2|\
\vec u_2\cdot(\vec u_5 - \vec u_4) + i\epsilon}\,
{ 1 \over 2|\vec\ell_4|}\,
{ 1 \over 2|\vec\ell_5|}
\nonumber\\
&= &
{ {\cal S}_2 \over 16\,|\vec\ell_4|^2\,|\vec\ell_5|^2}\,
{ 1 \over |\vec\ell_2|^3}\,
{ 1 \over 1 - (\vec u_2\cdot \vec u_4)^2}\,
{ 1 \over \vec u_2\cdot(\vec u_5 - \vec u_4) + i\epsilon}.
\end{eqnarray}
We proceed in this fashion to evaluate the contribution corresponding
to Fig.~\ref{fig:cuts8}(a2). Then we evaluate the contribution of
Fig.~\ref{fig:cuts8}(a3), but we find that this contribution is not
singular as $\vec \ell_2 \to 0$. Adding the three contributions, we
obtain the net integrand for the cut graph of  Fig.~\ref{fig:cuts}(a)
in the soft limit $\vec \ell_2 \to 0$:
\begin{equation}
{\cal G}_a \sim
{-  {\cal S}_2 \over 32\,|\vec\ell_4|^2\,|\vec\ell_5|^2}\,
{ 1 \over |\vec\ell_2|^3}\,
{ 1 \over 1 + \vec u_2 \cdot \vec u_4}\,
{ 1 \over 1 - \vec u_2 \cdot \vec u_5}\,
{ 2 - \vec u_2\cdot(\vec u_5 - \vec u_4)
\over \vec u_2\cdot(\vec u_5 - \vec u_4) + i\epsilon}.
\label{Gasoft}
\end{equation}

Some comments are in order here. First, we have included the leading
term, with a $1/|\vec\ell_2|^3$ singularity, and dropped less singular
terms. If we decompose the integration over $\vec\ell_2$ into $\int\!
d\Omega_2\,\int  |\vec\ell_2|^2 d|\vec\ell_2|$, then a
$1/|\vec\ell_2|^3$ singularity produces a logarithmic divergence in
the integration over $|\vec\ell_2|$. The less singular terms will lead
to a finite integration over  $|\vec\ell_2|$, although the integration
$\int\! d\Omega$  over the angles $\vec u_2$ can still be divergent.
There are, in fact, singularities in the angular integration. The
factor $1/[ 1 - \vec u_2 \cdot \vec u_5]$ is singular when $\vec
\ell_2$ is collinear with $\vec \ell_5$, while the factor $1/[ 1 + \vec
u_2 \cdot \vec u_4]$ is singular when $-\vec \ell_2$ is collinear with
$\vec \ell_4$. These singularities produce logarithmically divergent
integrations over $\vec u_2$. However, the analysis of the previous
subsection shows that the collinear singularities cancel among the
cuts of our graph. There is also a singularity on the plane
$\vec u_2\cdot(\vec u_5 - \vec u_4) = 0$. This is the scattering
singularity on the ellipse $|\vec\ell_1| + |\vec\ell_3| =
|\vec\ell_4| + |\vec\ell_5|$. This ellipse passes through the point
$\vec\ell_2 = 0$ and the plane tangent to the ellipse at this point is
the plane $\vec u_2\cdot(\vec u_5 - \vec u_4) = 0$. I will have more
to say about this singularity later. Here we note simply that it comes
with an $i\epsilon$ prescription, which has been preserved in
Eq.~(\ref{Gasoft}). 

We now consider the cut graph shown in Fig.~\ref{fig:cuts}(b). Again,
there are three contributions to consider, corresponding to the
diagrams $(b5)$, $(b2)$ and $(b4)$ in Fig.~\ref{fig:cuts8}. Adding the
three contributions, we obtain the net integrand for the cut graph of
Fig.~\ref{fig:cuts}(b) in the soft limit $\vec \ell_2 \to 0$:
\begin{equation}
{\cal G}_{b} \sim
{{\cal S}_2 \over 32\,|\vec\ell_4|^2\,|\vec\ell_5|^2}\,
{ 1 \over |\vec\ell_2|^3}\,
{ 1 \over 1 - \vec u_2 \cdot \vec u_4}\,
{ 1 \over 1 + \vec u_2 \cdot \vec u_5}\,
{ 2 + \vec u_2\cdot(\vec u_5 - \vec u_4)
\over \vec u_2\cdot(\vec u_5 - \vec u_4) + i\epsilon}.
\label{Gbsoft}
\end{equation}
As in Eq.~(\ref{Gasoft}), there are a scattering singularity and two
collinear singularities. However, the signs that indicate the location
of the collinear singularities are reversed compared to
Eq.~(\ref{Gbsoft}). If we add ${\cal G}_{a}$ and ${\cal G}_{b}$ we
obtain
\begin{equation}
{\cal G}_{a}+{\cal G}_{b} \sim
{- {\cal S}_2 \over 16\,|\vec\ell_4|^2\,|\vec\ell_5|^2}\,
{ 1 \over |\vec\ell_2|^3}\,
{ 1 + (\vec u_2\cdot\vec u_4)(\vec u_2\cdot\vec u_5)
\over 
[1 - (\vec u_2 \cdot \vec u_4)^2]
[1 - (\vec u_2 \cdot \vec u_5)^2]}.
\label{Gabsoft}
\end{equation}
Thus, the overall $1/|\vec\ell_2|^3$ singularity remains and the
collinear singularities remain, but the scattering singularities
cancel in the soft limit, $\vec \ell_2 \to 0$, between the two cuts
that leave virtual subgraphs.

There are two more cut graphs to consider. The graph shown in 
Fig.~\ref{fig:cuts}(c) gives
\begin{equation}
{\cal G}_c \sim
{{\cal S}_3 \over 32\,|\vec\ell_4|^2\,|\vec\ell_5|^2}\,
{ 1 \over |\vec\ell_2|^3}\,
{ 1 \over [1 + \vec u_2 \cdot \vec u_4]}\,
{ 1 \over [1 + \vec u_2 \cdot \vec u_5]}.
\label{Gcsoft}
\end{equation}
The graph shown in Fig.~\ref{fig:cuts}(d) gives
\begin{equation}
{\cal G}_d \sim
{{\cal S}_3 \over 32\,|\vec\ell_4|^2\,|\vec\ell_5|^2}\,
{ 1 \over |\vec\ell_2|^3}\,
{ 1 \over [1 - \vec u_2 \cdot \vec u_4]}\,
{ 1 \over [1 - \vec u_2 \cdot \vec u_5]}.
\label{Gdsoft}
\end{equation}
Adding these together, we find
\begin{equation}
{\cal G}_c+{\cal G}_d \sim
{{\cal S}_3 \over 16\,|\vec\ell_4|^2\,|\vec\ell_5|^2}\,
{ 1 \over |\vec\ell_2|^3}\,
{ 1 + (\vec u_2\cdot\vec u_4)(\vec u_2\cdot\vec u_5)
\over 
[1 - (\vec u_2 \cdot \vec u_4)^2]
[1 - (\vec u_2 \cdot \vec u_5)^2]}.
\label{Gcdsoft}
\end{equation}

We note that when we add the contributions of the cuts which leave
virtual subgraphs to the contributions of the cuts which have no
virtual subgraphs, the leading soft singularity cancels:
\begin{equation}
{\cal G}_a+{\cal G}_b+{\cal G}_c+{\cal G}_d \sim 0.
\label{Gsofttotal}
\end{equation}
That is, after cancellation, the overall singularity is at worst
proportional to $1/|\vec\ell_2|^2$. It is thus an integrable
singularity provided that all of the singularities of the angular
integration over $\vec u_2$ cause no problems.

The cancellation of the leading soft singularity is built into the
structure of Feynman diagrams, so that we do not have to do anything
special to make it happen. However, there is a certain subtlety in
arranging for the singularities in the angular integrations to be
convergent in a Monte Carlo integration. Thus, we will return to the
cancellation of the soft singularity after we have discussed contour
deformations in the following section.

\section{The scattering singularity and contour deformation}
\label{sec:deform}

Consider the contribution from Fig.~\ref{fig:cuts8}(a1), as given in
Eq.~(\ref{Gparts}). There is a factor
\begin{equation}
{1\over (|\vec\ell_1| - |\vec\ell_4| - |\vec\ell_5|)^2 
  -\vec\ell_3^{\,2} + i\epsilon}
=
{1\over (|\vec\ell_3| + |\vec\ell_4| + |\vec\ell_5| - |\vec\ell_1|)
 (|\vec\ell_4| + |\vec\ell_5| - |\vec\ell_1| - |\vec\ell_3|+
i\epsilon)},
\end{equation}
which has a singularity when $|\vec\ell_1| + |\vec\ell_3| =
|\vec\ell_4| + |\vec\ell_5|$. In an analysis using time-ordered
perturbation theory, the singular factor emerges from the energy
denominator associated with the intermediate state consisting of
partons 1 and 3, 
\begin{equation}
E_F - E(\vec\ell_2) + i\epsilon,
\label{denom2}
\end{equation}
where $E_F = |\vec\ell_4| + |\vec\ell_5|$ and
\begin{equation}
E(\vec\ell_2) 
=|\vec\ell_1| + |\vec\ell_3|
= 
|\vec\ell_4 + \vec\ell_2| 
+ |-\vec\ell_5 + \vec\ell_2|.
\end{equation}
Thus the singularity  appears when the momenta are right for
particles 1 and 3 to be on-shell and scatter to produce the final
state particles 4 and 5.

The contribution from  Fig.~\ref{fig:cuts8}(b4) has a scattering
singularity at the same place as that from the cut diagram (a1).
However, these singularities do not cancel in general because the
functions ${\cal S}_2(\vec\ell_4,\vec\ell_5)$ and ${\cal
S}_2(\vec\ell_1,\vec\ell_3)$ do not match. We thus have a problem if
we would like to perform the integration numerically.

We notice, however, that the singularity is protected by an
$i\epsilon$ prescription. The $i\epsilon$ in the denominator tells us
what to do in an analytic calculation and it also tells us what to do
in a numerical calculation: we need to deform the integration contour.

We are integrating over a loop momentum $\vec \ell_2$. Let us replace
$\vec \ell_2$ by a complex momentum $\vec \ell_{2,c} = \vec \ell_2 + i
\vec \kappa$, where $\vec\kappa$ is a function, which remains to be
determined, of $\vec \ell_2$. Then as we integrate over the real
vector $\vec \ell_2$, we are integrating over a contour in the space
of the complex vector $\vec \ell_{2,c}$. When we deform the original
contour $\vec \ell_{2,c} = \vec \ell_2$ to the new contour $\vec
\ell_{2,c} = \vec \ell_2 + i\vec \kappa$, the integral does not change
provided that we do not cross any points where the integrand is
singular and provided that we include a jacobian
\begin{equation}
{\cal J}(\vec \ell_2) = \det \left(\partial \ell_{2,c}^i\over 
\partial \ell_{2}^j\right).
\label{deformjacob}
\end{equation}
There are some subtleties associated with this; the relevant theorem
is proved in the Appendix.

We need to choose $\vec \kappa$ as a function of $\vec \ell_2$.
Consider first the direction of $\vec \kappa$. On the deformed
contour, the energy denominator (\ref{denom2}) has the form
\begin{equation}
E_F - E(\vec\ell_2 + i\vec\kappa) + i\epsilon.
\end{equation}
In order to fix the direction of deformation, it is useful to consider
what happens when we deform the contour just a little way from the
real $\vec \ell_2$ space. For small $\kappa$, we have
\begin{equation}
E(\vec\ell_2 + i\vec\kappa) \approx 
|\vec\ell_1| 
+ |\vec\ell_3|
+ i\vec\kappa \cdot \vec w,
\end{equation}
where
\begin{equation}
\vec w = { \vec\ell_1 \over |\vec\ell_1|}
+   { \vec\ell_3 \over |\vec\ell_3|}.
\label{wdef}
\end{equation}
Thus the energy denominator is
$
{  E_F - E(\vec\ell_2) 
- i \vec\kappa \cdot \vec w
+ i\epsilon}
$
for small $\vec\kappa$. In order to keep on the proper side of the
singularity, we want $\vec \kappa \cdot \vec  w$ to be negative. The
simplest way to insure this is to choose $\vec \kappa$ in the
direction of $-\vec w$. Thus we choose
\begin{equation}
\vec \kappa = - D(\vec \ell_2)\, \vec w,
\hskip 1 cm D(\vec \ell_2)\ge 0.
\end{equation}
Then the singular factor is approximately
\begin{equation}
{ 1 \over E_F - E(\vec\ell_2) 
+ i D(\vec \ell_2)\,\vec w^{\,2}}
\end{equation}
for a small deformation. For a larger deformation, it not so simple to
see that we stay on the correct side of the singularity, but
it is easy to check numerically.

The next question is how should we choose $D(\vec \ell_2)$? We want
$D$ not to be small when $\vec \ell_2$ is near the  surface
$E(\vec \ell_2) = E_F$ in order that the integrand not be large there.
We want $D(\vec \ell_2)$ not to grow as $\vec \ell_2^{\,2} \to \infty$
in order to satisfy the conditions for the theorem that deforming the
contour does not change value of the integral. Since there is no
reason to keep any finite contour deformation for large $\vec
\ell_2^{\,2}$, we will simply choose to have $D(\vec \ell_2) \to 0$ as
$\vec \ell_2^{\,2} \to \infty$.

There is another condition that $D(\vec \ell_2)$ should obey: it
should vanish at points where ${\cal G}$ has collinear and soft
singularities. To see why takes some discussion.

Consider the contributions from three parton cuts, for which there
is no virtual loop. For these contributions, we do not want to deform
the contours. This is because if any of the loop momenta were complex
then at least one of the momenta of the final state particles would be
complex. In principle, one could have complex momenta for final state
particles as long as the measurement functions ${\cal S}_n(\vec
k_1,\dots,\vec k_n)$ are analytic. However, I have in mind
applications in which the numerical integration program acts as a
subroutine that produces ``events'' with final state particle momenta
$\{\vec k_1,\dots,\vec k_n\}$ and weights computed by the subroutine.
Then the events could be the input to, for example, a Monte Carlo
program that generates parton showers and hadronization. Surely complex
momenta for the final state particles are not desirable.

Now recall that there is a cancellation among the contributions ${\cal
G}_C$ from different cuts $C$ at points where the ${\cal G}_C$ have
collinear and soft singularities. Evidently, if we deform the contour
for a contribution with a virtual graph but do not deform the contour
for the canceling contribution, then the cancellation can be spoiled. We
can avoid spoiling the cancellation if we make the contours match at
the singularity. That is, $D(\vec \ell_2)$ should vanish at the points
where the ${\cal G}_C$ have collinear and soft singularities.

We also need to determine how fast $D(\vec \ell_2)$ needs to approach
zero as $\vec \ell_2$ approaches a singularity. Since the integration
is in a multidimensional complex space, we need an analysis that makes
use of the multidimensional contour deformation theorem. This analysis
is given in the Appendix. Here, I present a simpler one dimensional
analysis that can serve to clarify the issue.

Consider the following toy integral,
\begin{equation}
I = \int_0^{x_{\rm max}} { d x \over x}\ \left\{
{f_V(x)\over x-1 + i\epsilon}  + f_R(x)
\right\}.
\end{equation}
Here the endpoint singularity at $x = 0$ plays the role of the
collinear or soft singularities. The function $f_V/(x-1 + i\epsilon)$
plays the role of the integrand for the contribution with a virtual
subgraph. In this contribution, there is a singularity at $x = 1$ that
comes with an $i\epsilon$ prescription. The function $f_R$
plays the role of the integrand for the contribution with no virtual
subgraph. We assume that $f_V(z)$ and $f_R(z)$ are analytic functions.
We also assume that $f_V(0) = f_R(0)$, so that the apparent
singularity at $x = 0$ cancels. 

Now the $i\epsilon$ prescription on the singularity at $x = 1$ tells
us that we can deform the integration contour into the upper half
plane, replacing $x$ by $z = x + i y(x)$ where $y(0) = y(x_{\rm max})
= 0$. Thus
\begin{equation}
I = \int_0^{x_{\rm max}}\! d x\
{  1 + i y'(x)\over x+iy(x)}\ 
\
\left\{
{f_V\left(x+iy(x)\right)\over x - 1 +iy(x)}  
+ f_R\left(x+iy(x)\right)
\right\}.
\end{equation}
Suppose, however, that we want to keep the contour for $f_R$ on the
real axis. Then we might hope that $I = \tilde I$, where
\begin{equation}
\tilde I = 
\lim_{x_{\rm min} \to 0}
\int_{x_{\rm min}}^{x_{\rm max}}\! d x\
\left\{
{ 1 + i y'(x)\over x+iy(x)}\ 
{f_V\left(x+iy(x)\right)\over x - 1 +iy(x)}  
+ {f_R\left(x\right)\over x}
\right\}.
\end{equation}
The difference is
\begin{equation}
\tilde I - I = 
\lim_{x_{\rm min} \to 0}
\int_{x_{\rm min}}^{x_{\rm max}}\! d x\
\left\{  
{f_R\left(x\right)\over x}
- [1 + i y'(x)]
{f_R\left(x+iy(x)\right)\over x+iy(x)}\ 
\right\}.
\end{equation}
If we note that $[f_R(z) - f_R(0)]/z$ is an analytic function even at
$z = 0$ and that the integral of an analytic function around a closed
contour vanishes, we have
\begin{equation}
0 = 
\lim_{x_{\rm min} \to 0}
\int_{x_{\rm min}}^{x_{\rm max}}\! d x\
\left\{  
{f_R\left(x\right) - f_R\left(0\right)\over x}
- [1 + i y'(x)]
{f_R\left(x+iy(x)\right) - f_R\left(0\right)\over x+iy(x)}\ 
\right\}.
\end{equation}
Subtracting these and performing the integral, we have
\begin{equation}
\tilde I - I = f_R(0)\
\lim_{x_{\rm min} \to 0}
\int_{x_{\rm min}}^{x_{\rm max}}\! d x\
\left\{  
{1\over x}
- 
{1 + i y'(x)\over x+iy(x)}\ 
\right\}
= f_R(0)\
\lim_{x_{\rm min} \to 0}
\log\left(
1 + i\, {y(x_{\rm min})\over x_{\rm min}}
\right).
\end{equation}

We can draw two conclusions. First, as long as $y(x)\to 0$ at least
as fast as $x^1$ as $x \to 0$, we will realize the cancellation of the
$x \to 0$ singularity and obtain a finite value for $\tilde I$.
Second, if we choose $y(x) \propto x$ as $x \to 0$, $\tilde I$ will be
finite, but it will not be equal to the correct result $I$. In order
to get a result $\tilde I$ that is not only finite but also correct,
we need $y(x)/x \to 0$ as $x\to 0$. A convenient choice is
$y(x) \propto x^2$ as $x\to 0$.

We conclude from the multidimensional extension of this analysis, given
in the Appendix, that as $\vec\ell_2$ approaches a singularity,
$D(\vec\ell_2)$ should approach zero quadratically with the distance to
the singularity.

We now use the qualitative criteria just developed to give a specific
choice of deformation.  We have chosen
\begin{equation}
\vec\ell_{2,c} = \vec\ell_2 - i D(\vec\ell_2)\vec w,
\label{deform}
\end{equation}
where $\vec w$, Eq.~(\ref{wdef}), specifies the direction of
deformation. We now specify a deformation function $D(\vec\ell_2)$
that satisfies our criteria. We write $D$ in the form
\begin{equation}
D = C\,G.
\label{DisCG}
\end{equation}
The factor $C$ is designed to insure that the deformation vanishes
quadratically near the collinear and soft singularities. The factor
$G$ is designed to turn the deformation off for large $\vec\ell_2$.
These factors are explained below and are defined precisely in
Eqs.~(\ref{Cdef}) and (\ref{Gdef}) below.

First, we discuss the factor $C$. We want the deformation to vanish
at the line $\vec\ell_2 = - x \vec\ell_4$ with $0 \le x \le
1$, where the amplitude has a collinear singularity. (Since
$\vec\ell_4 = \vec\ell_1 -\vec\ell_2$, this line is also $\vec\ell_1 =
-\lambda \vec\ell_2$ with $0 < \lambda < \infty$.) Define
\begin{equation}
d_{12} = 
{\left|  |\vec\ell_2| \vec\ell_1 + |\vec\ell_1| \vec\ell_2\right| 
\over 
|\vec\ell_1 - \vec\ell_2|}
=
{ \left| |\vec\ell_2| \vec\ell_1 + |\vec\ell_1| \vec\ell_2\right|
\over 
 |\vec\ell_4|}.
\label{d12def}
\end{equation}
This function is zero on the line $\vec\ell_2 = -x \vec\ell_4$
with $0\le x \le 1$, and furthermore, it vanishes linearly 
as $\vec\ell_2$ approaches this line.  Similarly,  we want the
deformation to vanish on the line $\vec\ell_2 = x \vec\ell_5$
with $0\le x \le 1$, where the amplitude has its other collinear
singularity. The function $d_{23}$, where
\begin{equation}
d_{23} = 
{ \left||\vec\ell_3| \vec\ell_2 + |\vec\ell_2|\vec\ell_3\right| 
\over 
|\vec\ell_2 - \vec\ell_3|}
=
{ \left| |\vec\ell_3| \vec\ell_2 + |\vec\ell_2| \vec\ell_3\right|
\over  |\vec\ell_5|},
\label{d23def}
\end{equation}
vanishes linearly as $\vec\ell_2$ approaches this line. (To see this,
use $\vec\ell_5 = \vec\ell_2 - \vec\ell_3$.)  Let
\begin{equation}
d = \min(d_{12},d_{23}).
\label{ddef}
\end{equation}
Then $d$ vanishes linearly with the distance to either of the
collinear singularities. It also vanishes linearly with the distance to
the soft singularity at $\vec \ell_2 = 0$.

Now, we have seen that the deformation should vanish quadratically with
the distance to any of the singularities. We can achieve this by
letting
\begin{equation}
C(d^2) = { \alpha\, d^2 \over 
1 + 4 \beta d^2/(|\ell_4| + |\ell_5| + |\vec q|)^2},
\label{Cdef}
\end{equation}
where $\alpha$ and $\beta$ are adjustable dimensionless parameters.
Note that, for large $d$, $C(d^2)$ approaches a constant.

Next, we discuss the factor $G$. We want to ensure that the contour
deformation vanishes for large $\vec \ell_2$. Let us define
\begin{equation}
a = |\vec\ell_1| + |\vec\ell_3| - |\vec q|
\label{adef}
\end{equation}
and
\begin{equation}
A = |\vec\ell_4| + |\vec\ell_5| - |\vec q|.
\label{Adef}
\end{equation}
Then the singularity that we are avoiding by means of contour
deformation is at $a = A$.  We can turn the deformation off for $a \gg
A$ by setting
\begin{equation}
G(a) = { 1 \over A + \gamma a},
\label{Gdef}
\end{equation}
where $\gamma$ is an adjustable dimensionless parameter. 

There is a subsidiary reason for this choice. At the singularity, $G =
1/[(1+\gamma)A]$. The factor $1/A$ serves to enhance the deformation
in the case that $\vec \ell_4$ and $\vec\ell_5$ are nearly collinear,
in which case $d$ is small on the ellipse $a = A$ and the deformation
would otherwise be too small.

The reader will note that, while there is a certain uniqueness in
defining the direction of the deformation in Eq.~(\ref{deform}) to be
given by the vector $\vec w$, Eq.~(\ref{wdef}), the normalization $D =
C\,G$ with $C$ and $G$ given in  Eqs.~(\ref{Cdef}) and
(\ref{Gdef}) is rather {\it ad hoc}. Within the requirements that the
deformation should vanish quadratically at the collinear and soft
singularities and should vanish for large $\vec \ell_2$, many other
choices would be possible. The choice given here is used in the
current version of the code \cite{beowulf}. Surely there is some other
choice that is better.

\section{The Monte Carlo integration}
\label{sec:MonteCarlo}

After the contour deformations, we have an integral of the form
\begin{equation}
{\cal I} 
= \int\! d \ell\,
\sum_C\,
{\cal J}(C;\ell)\
g(C;\ell + i\kappa(C;\ell)),
\end{equation}
where we use $\ell$ for the loop momenta collectively, $\ell = \{\vec
\ell_2, \vec\ell_4\}$. The index $C$ labels the cut, $a$, $b$, $c$, or
$d$ in Fig.~\ref{fig:cuts}. There is a contour deformation that
depends on the cut, as specified by $\kappa(C;\ell)$, and there is a
corresponding jacobian ${\cal J}(C;\ell)$, Eq.~(\ref{deformjacob}). 
Define
\begin{equation}
f(\ell) = \Re\left\{
\sum_C {\cal J}(C;\ell)\
g(C;\ell + i\kappa(C;\ell)) \right\}.
\end{equation}
We know that $\cal I$ is real, so 
\begin{equation}
{\cal I} = \int\! d\ell\,f(\ell).
\end{equation}

To perform the integration, we use the Monte Carlo method. We choose
points $\ell$ with a density $\rho(\ell)$,
with
\begin{equation}
\int\!d\ell\, \rho(\ell) = 1.
\end{equation}
After choosing $N$ points $\ell_1,\dots,\ell_N$, we have an estimate for
the integral:
\begin{equation}
{\cal I}  \approx {\cal I}_N =
{1\over N}\sum_i{f(\ell_i) \over \rho(\ell_i)}.
\end{equation}
This is an approximation for the integral in the sense that if we
repeat the procedure a lot of times the expectation value for 
${\cal I}_N$ is 
\begin{equation}
\langle {\cal I}_N \rangle = {\cal I}.
\end{equation}
The expected r.m.s.\ error is ${\cal E}$, where
\begin{equation}
{\cal E}^2 = \langle \left( {\cal I}_N - \cal I\right)^2\rangle
= { 1 \over N}
\int\!d\ell\,{ f(\ell)^2 \over \rho(\ell)}
- {{\cal I}^2 \over N}.
\end{equation}
One can rewrite this as
\begin{equation}
{\cal E}^2 =
{ 1 \over N}\int\!d\ell\,\rho(\ell)
\left( {|f(\ell)| \over \rho(\ell)} - \tilde {\cal I}
\right)^2
+ { \tilde {\cal I}^2 - {\cal I}^2 \over N},
\end{equation}
where
\begin{equation}
\tilde{\cal I} = \int\!d\ell\,|f(\ell)|.
\end{equation}
We see, first of all, that the expected error decreases proportionally
to $1/\sqrt N$. Second, we see that the ideal choice of $\rho(\ell)$
would be $\rho(\ell) = |f(\ell)|/\tilde {\cal I}$.

Of course, it is not possible to choose $\rho$ in this way. But we know
that $|f|$ has singularities at places where propagator momenta vanish
and we know the structure of these singularities. We are not really
able to choose $\rho$ so that ${|f(\ell)| / \rho(\ell)}$ is a
constant, but at least we can choose it so that ${|f(\ell)| /
\rho(\ell)}$ is not singular at the singularities of $|f(\ell)|$.

Note that it is easy to combine methods for choosing Monte Carlo
points. Suppose that we have a recipe for choosing points with a
density $\rho_1$ that is singular when one propagator momentum
vanishes, a  recipe for choosing points with a density
$\rho_2$ that is singular when another propagator momentum vanishes,
and in general recipes for choosing points with densities $\rho_i$
with several goals in mind. Then we can devote a fraction $\lambda_i$
of the points to the choice with density $\rho_i$ and obtain a net
density
\begin{equation}
\rho(\ell) = \sum_i \lambda_i\, \rho_i(\ell).
\end{equation}

\subsection{The density near where a propagator momentum vanishes}

Let $\vec \ell_J$ be the momentum of one of the propagators in our
graph. We have seen that when particle $J$ appears in the final state,
there is a factor $1/|\vec \ell_J|$ in the integrand. When  propagator
$J$ is part of a virtual loop, the contribution corresponding to
this propagator being put on shell also contains a factor $1/|\vec
\ell_J|$. Thus there is a singularity $1/|\vec \ell_J|$ for every
propagator in the graph.

The analysis given in the introduction to this section indicates that
for each propagator $J$ one of the terms $\rho_i$ in the density
function should have a singularity that is at least as strong as
\begin{equation}
\rho_i(\ell) \propto 1/|\vec\ell_J|
\end{equation}
as $\vec \ell_J \to 0$. It is, of course, easy to choose points with a
density proportional to $1/|\vec\ell_J|^A$ as $\vec \ell_J \to 0$ as
long as $A < 3$. (The limitation on $A$ arises because for $A \ge 3$ we
would have $\int d\vec\ell_J\, \rho(\ell) = \infty$.) Thus it is easy
to arrange that the density of points has the requisite singularities.
Specifically, we can choose $\vec\ell_J$ with the density
\begin{equation}
\tilde\rho(\vec\ell_J) = {1\over 2\pi K_0^3}\
{1 \over \left[1+\left(|\vec\ell_J|/K_0\right)^2\right]^2}\
{K_0 \over |\vec \ell_J|},
\label{mildrho}
\end{equation}
where $K_0$ is a momentum scale determined by the other, previously
chosen, loop momenta.

The singularity when $\vec \ell_J \to 0$ can be more severe than
$1/|\vec\ell_J|$, depending on the structure of the graph. Consider
first the cases $J = 1,3,4,5$. Here, the singularities for particular
cuts, as given in Eq.~(\ref{Gparts}), are $1/|\vec\ell_J|^2$. However,
there is a cancellation after one sums over cuts (as for the
singularity for $\vec \ell_2 \to 0$), leaving a singularity
$1/|\vec\ell_J|$. 

For $J = 2$ there is a severe singularity of the form
$1/|\vec\ell_J|^3$ for particular contributions to Eq.~(\ref{Gparts}).
A $1/|\vec\ell_2|^3$ singularity would not be integrable, but, as we
have seen in detail, there is a cancellation among the contributions so
that only a $1/|\vec\ell_2|^2$ singularity is left. However, it will
not do to simply chose $\rho_i(\ell) \propto 1/|\vec\ell_2|^2$ because
there is also a singularity in the space of the angles of $\vec\ell_2$.
It is to this subject that we now turn. 

\subsection{The soft parton singularity}

When two partons can scatter by exchanging a parton before they enter
the final state, there is a severe singularity as the momentum of
the exchanged parton goes to zero. For our graph, this happens for 
$\vec\ell_2 \to 0$. In this subsection, we consider the behavior of
the integrand for small $\vec\ell_2$ as a function of its
magnitude $\vec\ell_2$ and of its direction $\vec u_2 = \vec
\ell_2/|\vec \ell_2|$.

The singularity for individual cuts, as given in  Eq.~(\ref{Gparts}),
is of the form $1/|\vec\ell_2|^3$ when we let $|\vec\ell_2| \to 0$ with
$\vec u_2$ held fixed. This singularity is not integrable. However, as
we have seen, the leading term cancels when we sum over cuts, leaving a
$1/|\vec\ell_2|^2$ singularity for $|\vec\ell_2| \to 0$ with $\vec u_2$
fixed.

Let us now recall from Eq.~(\ref{Gasoft}) that, before we deform the
integration contour, the contribution for small $\vec \ell_2$ from the
cut $a$ of Fig.~\ref{fig:cuts} has, in addition to a factor
$1/|\vec\ell_2|^3$, a factor $1/[\vec u_2\cdot(\vec u_5 - \vec u_4) +
i \epsilon]$. That is, there is a singularity on a surface in the
space of $\vec \ell_2$ whose tangent plane is the plane perpendicular
to $\vec u_5 - \vec u_4$. We have avoided this singularity by
deforming the integration contour. However, the deformation vanishes
as $\vec\ell_2 \to 0$. Thus we must face the question of what happens
to the cancellation near the soft parton singularity when the contour
deformation is taken into account.

First, let us recall from Eq.~(\ref{deform}) that for cut $a$ in 
Fig.~\ref{fig:cuts} the deformation has the form 
\begin{equation}
\vec\ell_{2,c} = \vec\ell_2 - i D(\vec\ell_2)\,\vec w,
\end{equation}
where $\vec w = \vec u_1 + \vec u_3$. For $\vec \ell_2 \to 0$,
\begin{equation}
\vec w \sim \vec u_4 - \vec u_5 ,
\end{equation}
while $D$ has the form 
\begin{equation}
D(\vec \ell_2) \sim \vec \ell_2^{\,2}\ \tilde D(\vec u_2).
\end{equation}
Here $\tilde D$ vanishes for $\vec u_2 = - \vec u_4$ and for $\vec u_2
= \vec u_5$ and is positive elsewhere. Thus
\begin{equation}
\vec\ell_{2,c} = |\vec\ell_2|\left(\vec u_2  
+ i |\vec\ell_2|\tilde D(\vec u_2)\, (\vec u_5 - \vec u_4)
+ {\cal O}(\vec\ell_2^{\,2})
\right).
\label{ell2c}
\end{equation}

Substituting $\ell_{2,c}$ as given above for $\vec\ell_2$ in
Eq.~(\ref{Gasoft}), we obtain an expression for the contribution from
cut $a$ to the integrand on the deformed contour near the soft
singularity:
\begin{equation}
{\cal G}_a \sim
{-  {\cal S}_2 \over 32\,|\vec\ell_4|^2\,|\vec\ell_5|^2}\,
{ 1 \over |\vec\ell_2|^3}\,
{ 1 \over 1 + \vec u_2 \cdot \vec u_4}\,
{ 1 \over 1 - \vec u_2 \cdot \vec u_5}\,
{ 2 - \vec u_2\cdot(\vec u_5 - \vec u_4)
\over \vec u_2\cdot(\vec u_5 - \vec u_4) 
+ i |\vec\ell_2|\tilde D(\vec u_2)\, (\vec u_5 - \vec u_4)^2}.
\label{Gasoftbis}
\end{equation}
There are two cases to consider. First, when $|\vec\ell_2| \to 0$ with
$\vec u_2$ fixed, we can drop the second term in the last denominator.
Then ${\cal G}_a \sim h(\vec u_2)/ |\vec\ell_2|^3$, where the
function $h(\vec u_2)$ is the same as on the undeformed contour. As we
have seen, the leading $1/ |\vec\ell_2|^3$ terms cancel when one sums
over cuts. Thus, as noted earlier, the net integrand behaves like 
\begin{equation}
{\cal G} \sim h_{\rm tot}(\vec u_2)/|\vec\ell_2|^2
\end{equation}
when $|\vec\ell_2| \to 0$ with $\vec u_2$ fixed.

The second case is more interesting. Consider $|\vec\ell_2| \to 0$ and
$\vec u_2\cdot(\vec u_5 - \vec u_4) \to 0$ with
$\vec u_2\cdot(\vec u_5 - \vec u_4)/|\vec\ell_2|$ fixed.
Then ${\cal G}_a$ is more singular, ${\cal G}_a \propto
1/|\vec\ell_2|^4$. To see what happens in this region, we analyze the
contribution from cut $b$ in Fig.~\ref{fig:cuts} in the same fashion.
The contour deformation for cut $b$ is different from that for cut
$a$, but the deformations match at leading order as $|\vec\ell_2| \to
0$. (This is an important feature of the choice of contour
deformations.) Thus we can use Eq.~(\ref{ell2c}) in Eq.~(\ref{Gbsoft})
to obtain
\begin{equation}
{\cal G}_{b} \sim
{{\cal S}_2 \over 32\,|\vec\ell_4|^2\,|\vec\ell_5|^2}\,
{ 1 \over |\vec\ell_2|^3}\,
{ 1 \over 1 - \vec u_2 \cdot \vec u_4}\,
{ 1 \over 1 + \vec u_2 \cdot \vec u_5}\,
{ 2 + \vec u_2\cdot(\vec u_5 - \vec u_4)
\over \vec u_2\cdot(\vec u_5 - \vec u_4) 
+ i |\vec\ell_2|\tilde D(\vec u_2)\, (\vec u_5 - \vec u_4)^2}.
\label{Gbsoftbis}
\end{equation}
We see that ${\cal G}_{b}$ is also proportional to $1/|\vec\ell_2|^4$
in the problematic region. However, since $u_2 \cdot u_4 \sim u_2 \cdot
u_5$ in this region, the leading $1/|\vec\ell_2|^4$ behavior cancels
when we add ${\cal G}_{b}$ to ${\cal G}_{a}$. We are left with the
next term, proportional to $1/|\vec\ell_2|^3$.

For the two remaining cuts there is no contour deformation. The
contributions from these cuts are each proportional to
$1/|\vec\ell_2|^3$. Calculation shows that there is no further
cancellation. Thus the net behavior of the integrand is
\begin{equation}
{\cal G} \propto 1/|\vec\ell_2|^3
\end{equation}
when $|\vec\ell_2| \to 0$ and
$\vec u_2\cdot(\vec u_5 - \vec u_4) \to 0$ with
$\vec u_2\cdot(\vec u_5 - \vec u_4)/|\vec\ell_2|$ fixed.

\subsection{Density near a soft parton singularity}
\label{subsec:nearsoft}

According to the analysis at the beginning of this section, we should
choose a density of integration points that has a singularity that
is at least as strong as that of $|{\cal G}|$ near the soft singularity
at $\vec\ell_2 \to 0$. Thus we should choose one of the $\rho_i$ so
that
\begin{eqnarray}
\rho_i(\ell) &\propto& 
{ 1 \over |\vec\ell_2|^p },
\hskip 1 cm
|\vec\ell_2| \to 0,\ \vec u_2 \ \ {\rm fixed},
\nonumber\\
\rho_i(\ell) &\propto& 
{ 1 \over |\vec\ell_2|^{p+1} },
\hskip 1 cm
|\vec\ell_2| \to 0,\ 
{\vec u_2 \cdot(\vec u_5 - \vec u_4) \over |\vec\ell_2|}\ \
{\rm fixed},
\end{eqnarray}
with $p\ge 2$.

Specifically, having chosen $\vec\ell_4$ we can choose the remaining
loop momentum $\vec\ell_2$ with the density
\begin{equation}
\tilde\rho(\vec\ell_2) =
{1\over 2\pi K_0^3 }\
{1\over \left[
1 + \left(|\vec\ell_2|/ K_0\right)^{(3-p)}
\right]^{(5-p)/(3-p)} }
\left({K_0 \over |\vec\ell_2|}\right)^{p}
{1\over \Gamma \sqrt{\cos^2(\theta) + \vec\ell_2^{\,2}/K_0^2}}.
\label{strongrho}
\end{equation}
Here $K_0$ is a momentum scale determined by $\vec\ell_4$, $\theta$ is
the angle between $\vec \ell_2$ and $(\vec u_5 - \vec u_4)$, and
\begin{equation}
\sinh (\Gamma) = K_0/|\vec\ell_2|.
\end{equation}
It is easy to choose points with this density by first choosing
$|\vec\ell_2|$, then choosing $\cos(\theta)$, and finally choosing the
corresponding azimuthal angle $\phi$ with a uniform density. 
Accounting for the fact that $\Gamma \propto \log(|\vec\ell_2|)$ for
$\vec\ell_2 \to 0$, we see that $\tilde\rho$ will have a singularity
stronger than that of ${\cal G}$ provided that $p > 2$. We will see how
this works in a numerical example in the next section.

\section{Numerical Example}
\label{sec:example}

In this section, I illustrate the principles developed above by means
of a particular example. We consider the integral in
Eq.~(\ref{theintegral}). We hold $\vec\ell_4$ fixed and consider the
integrand as a function of $\vec \ell_2$. In order to simplify the labelling, I define
\begin{equation}
\vec \ell_2 \equiv \vec \ell.
\end{equation}
There is a contribution for each cut $C$, with $C = a,b,c,\ {\rm
or}\ d$. For each contribution from a cut $C$ in which there is a
virtual loop, we want to deform the integration contour as discussed in
Sec.~\ref{sec:deform}. Thus $\vec \ell$ gets replaced by a complex
vector $\vec\ell_{c} = \vec\ell + i\vec\kappa_C$ and we need to
supply a jacobian ${\cal J}_C(\vec\ell)$, Eq.~(\ref{deformjacob}).
Then the integration over $\vec\ell$ has the form
\begin{equation}
\int\!d\vec\ell\, \sum_C {\cal J}_C(\vec\ell)\
{\cal G}_C(\vec\ell).
\end{equation}
The functions ${\cal G}_C$ are the analytic continuations to the
deformed contours of the functions given in  Eq.~(\ref{Gparts}). As
discussed in Sec.~\ref{sec:MonteCarlo}, the quantity that is relevant
for the convergence of the Monte Carlo integration is the integrand
divided by the density of points chosen for the integration. In this
section, I consider only the integration over $\vec\ell$, so I
discuss a choice for the density of integration points
$\rho(\vec\ell)$ at a fixed $\vec\ell_4$ and display plots of
the functions
\begin{equation}
F_C(\vec\ell) \equiv
{1\over\rho(\vec\ell)}\
{\cal J}_C(\vec\ell)\
{\cal G}_C(\vec\ell)
\end{equation}
and $F(\vec\ell) = \sum_C F_C(\vec\ell)$, as well as plots of the
deformation and the density.

For the numerical examples, I choose
\begin{equation}
\vec q = (3,-0.5,0)
\end{equation}
and then take $\vec\ell_4$ at the point
\begin{equation}
\vec \ell_4 = (2,-1,0).
\end{equation}
Since $\vec \ell_5 = \vec q - \vec \ell_4$ we have
\begin{equation}
\vec \ell_5 = (1,0.5,0).
\end{equation}
The singularities of the functions ${\cal G}_C(\vec \ell)$ lie in the
plane of $\vec\ell_4$ and $\vec\ell_5$, that is the $\ell_{z} = 0$
plane. In the plots, I choose $\ell_{z} = 0$, so that we see the
effect of the singularities. I plot $|\vec\kappa_a|$, $|\vec\kappa_b|$,
$\rho$, $F_a$, $F_b$, $F_c + F_d$ and $F$ as functions of $\ell_{x}$
and $\ell_{y}$ in the domain $-2.5 < \ell_{x} < 1.0$ and $-1.0 <
\ell_{y} < 2.0$.

\begin{figure}[htbp]
\centerline{\DESepsf(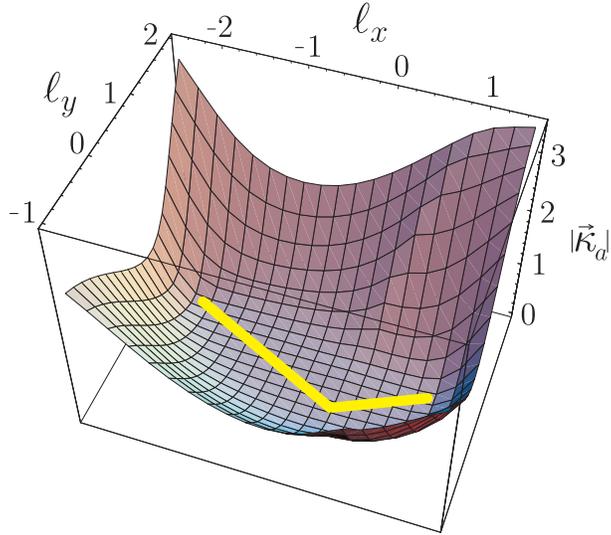 width 8 cm)}
\medskip
\caption{Contour deformation for cut $a$. The absolute value of
the imaginary part $\vec\kappa_a$ of $\vec\ell_{c}$ for cut $a$ is
plotted against $\ell_{x}$ and $\ell_{y}$ at
$\ell_{z} = 0$. The deformation $\vec\kappa_a$ vanishes at the
two collinear singularities for this cut, which are indicated by lines
superimposed on the graph.}
\label{fig:deformA}
\end{figure}

Consider first the contour deformation for cut $a$, $\vec\ell \to
\vec\ell_{c} = \vec\ell + i\vec\kappa_a$. I take $\vec\kappa_a = -
D\,\vec w$ as given in Eqs.~(\ref{deform}-\ref{Gdef}) with $\alpha =
\beta = \gamma = 1$. In Fig.~\ref{fig:deformA}, I show a graph of
$|\vec\kappa_a|$ versus $\ell_{x}$ and $\ell_{y}$. We see
that the deformation is not small. I also display in the figure the
lines $\vec\ell = -x\vec\ell_4$ with $0<x<1$ and $\vec\ell = 
x\vec\ell_5$ with $0<x<1$, where the collinear singularities for cut
$a$ are located.  We see that, as desired, the deformation vanishes
quadratically as $\vec\ell$ approaches these lines.

\begin{figure}[htbp]
\centerline{\DESepsf(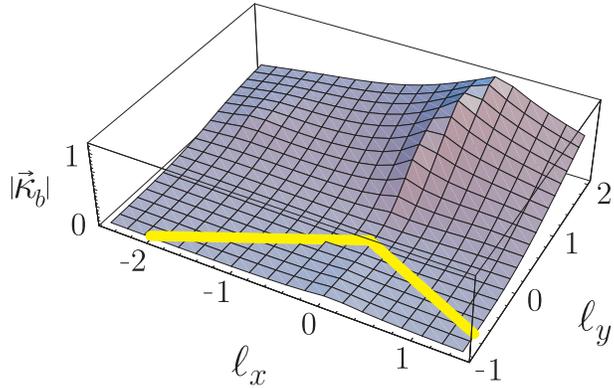 width 8 cm)}
\medskip
\caption{Contour deformation for cut $b$. The absolute value of
the imaginary part $\vec\kappa_b$ of $\vec\ell_{c}$ for cut $b$ is
plotted against $\ell_{x}$ and $\ell_{y}$ at
$\ell_{z} = 0$. The deformation $\vec\kappa_b$ vanishes at the
two collinear singularities for this cut, which are indicated by lines
superimposed on the graph.}
\label{fig:deformB}
\end{figure}

There is a different contour deformation for cut $b$. The same
formulas apply as for cut $a$  with the replacements $\vec\ell_4
\leftrightarrow \vec\ell_1$, $\vec\ell_5 \leftrightarrow -\vec\ell_3$,
$\vec\ell \leftrightarrow -\vec\ell$ and with the sign of
$\vec\kappa$ reversed. I show a graph of $|\vec\kappa_b|$ versus
$\ell_{x}$ and $\ell_{y}$ in Fig.~\ref{fig:deformB}.
(This figure does not look like Fig.~\ref{fig:deformA} because
$\vec\ell$ varies with $\vec\ell_4$ held fixed, not with $\vec\ell_1$
held fixed as would be needed if we applied the replacement 
$\vec\ell_4 \leftrightarrow \vec\ell_1$ to Fig.~\ref{fig:deformA}.)
I also display in the figure the lines $\vec\ell = \lambda\vec\ell_4$
with $0<\lambda$ and $\vec\ell =  -\lambda\vec\ell_5$ with
$0<\lambda$, where the collinear singularities for cut $b$ are
located. The deformation vanishes quadratically as $\vec\ell$
approaches these lines.

The jacobian  functions ${\cal J}_a(\vec\ell)$ and ${\cal
J}_b(\vec\ell)$ associated with the contour deformations are quite
unremarkable, so I omit showing them.

\begin{figure}[htbp]
\centerline{\DESepsf(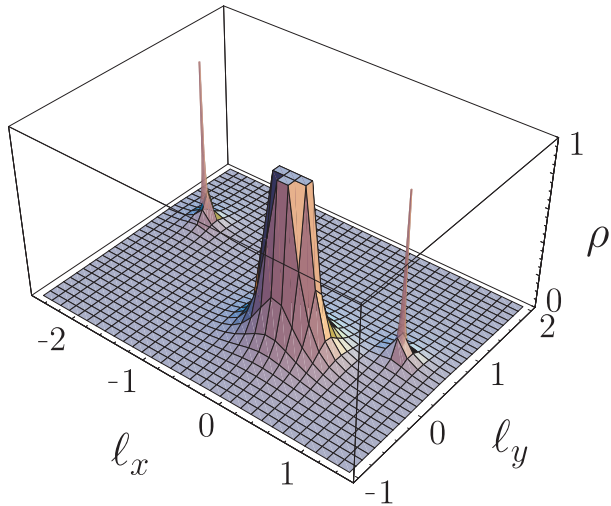 width 8 cm)}
\medskip
\caption{Density of integration points. The density has three
singularities. Only values $\rho < 1$ are shown.}
\label{fig:density}
\end{figure}

\begin{figure}[htbp]
\centerline{\DESepsf(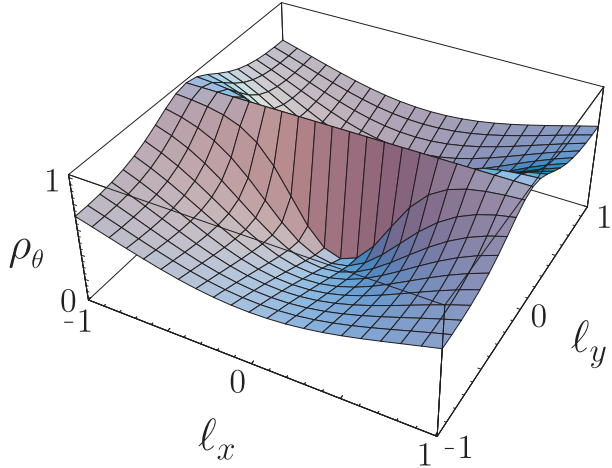 width 8 cm)}
\medskip
\caption{Angle dependent factor for the density $\rho_2$ of
integration points associated with the soft singularity. For small
$|\vec\ell|$, points are concentrated near $\theta = 0$. }
\label{fig:rhotheta}
\end{figure}

Consider next the density of integration points. I choose
\begin{equation}
\rho(\vec\ell) = 
0.2\,\rho_1(\vec\ell_1) +
0.6\,\rho_2(\vec\ell) +
0.2\,\rho_3(\vec\ell_3),
\end{equation}
as shown in Fig.~\ref{fig:density}. The function $\rho_1$ has a mild
singularity as $\vec\ell_1 \to 0$ and is given by Eq.~(\ref{mildrho})
with $\vec\ell_1 = \vec\ell_4 - \vec\ell$ and with $K_0$ set equal to
2. The function $\rho_3$ has a mild singularity as $\vec\ell_3 \to 0$;
I use the same functional form with $\vec\ell_3 = \vec\ell -
\vec\ell_5$. For $\rho_2$, I use the function given in
Eq.~(\ref{strongrho}) with $K_0 =  2$ and with the power $p$ taken as
$p = 2.2$. Then $\rho_2$ has a strong
$1/[|\vec\ell|^{2.2}\log(|\vec\ell|)]$ singularity as we approach
the $\vec\ell = 0$. Furthermore, the density of points is largest
near the plane $\ell_{y} = 0$, the plane that is tangent at
$\vec\ell = 0$ to the ellipsoidal surface that (if we turn off the
deformation) contains the scattering singularity. In order to display
the dependence of $\rho$ on angle near $\vec\ell = 0$, I plot in
Fig.~\ref{fig:rhotheta} the angle dependent factor in $\rho_2$,
namely the factor
\begin{equation}
{|\vec\ell|/K_0 \over \sqrt{\cos^2(\theta) +
\vec\ell^{\,2}/K_0^2}}
\end{equation}
in Eq.~(\ref{strongrho}), in a region near $\vec\ell = 0$. Here
$\cos(\theta) = \ell_{y}/|\vec\ell|$. We see that the density of
integration points is heavily concentrated very near the plane
$\ell_{y} = 0$ when $|\vec\ell|$ is small.

\begin{figure}[htbp]
\centerline{\DESepsf(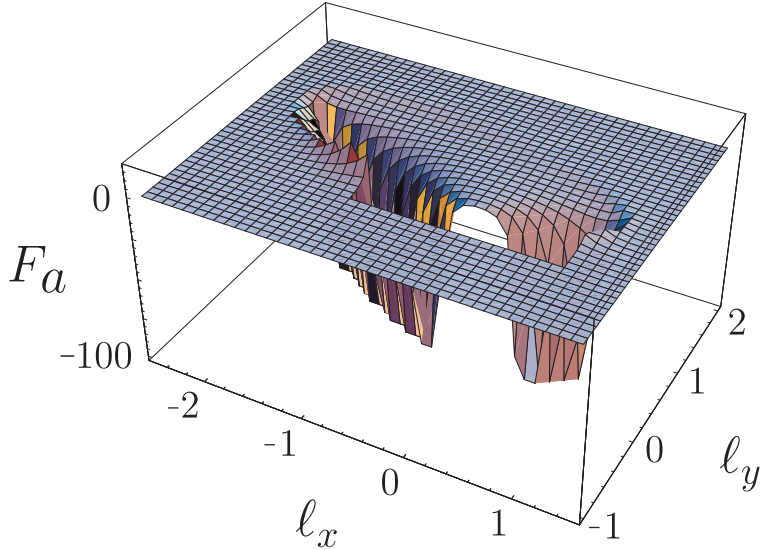 width 10 cm)}
\medskip
\caption{Contribution to $F$ from cut $a$. The domain
$-1<\ell_{x}<1$,  $-0.3<\ell_{y}<0.3$, which contains
the soft singularity, has been removed from the plot in order
to make the collinear singularities visible.  The function $F_a$ is
singular along lines from $(-2,1)$ to $(0,0)$ and from $(0,0)$ to
$(1,0.5)$. Only values $F_a>-100$ are shown. }
\label{fig:Fa}
\end{figure}

We are now ready to look at the contribution
$
F_a =
{\cal J}_a\
{\cal G}_a
/\rho
$
to $F$ from cut $a$. This function is displayed in Fig.~\ref{fig:Fa}
with a small rectangle near $\vec\ell = 0$ removed from the graph.
We see the two collinear singularities, at $\vec\ell = - x
\vec\ell_4$ and at $\vec\ell = x \vec\ell_5$ with $0<x<1$. As
$\vec\ell$ approaches one of these singularities, $F_a$ approaches
$-\infty$. 

In the standard method for calculating ${\cal I}$, we would perform
the integration over $\vec\ell$ analytically for the contribution
from cut $a$. Because of the singularities, the integration is
divergent. However, we can get a finite answer if we regulate the
integral by working in $3-2\epsilon$ spatial dimensions. Then the
result contains terms proportional to $1/\epsilon^2$ and $1/\epsilon$
as well as a remainder that is finite as $\epsilon \to 0$. 

\begin{figure}[htbp]
\centerline{\DESepsf(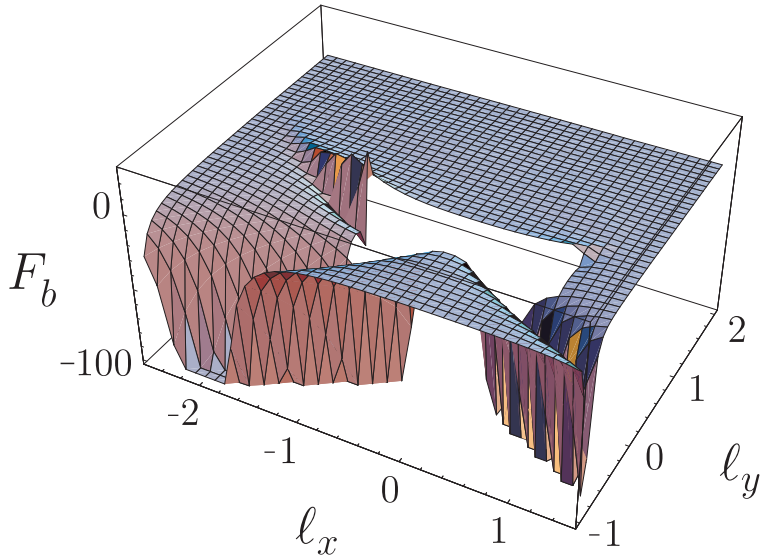 width 10 cm)}
\medskip
\caption{Contribution to $F$ from cut $b$. The same rectangular domain
as in Fig.~\ref{fig:Fa} has been removed from the plot.  The function
$F_b$ is singular along lines that extend from $(0,0)$ to infinity in
the directions $(-2,-1)$ and $(2,-1)$. Only values $F_b>-100$ are
shown. }
\label{fig:Fb}
\end{figure}

What about the contribution to $F$ from cut $b$, the other cut for
which there is a virtual subgraph? This function is displayed in
Fig.~\ref{fig:Fb} with the same small rectangle near $\vec\ell =
0$ removed from the graph. We see the two collinear singularities, at
$\vec\ell = \lambda \vec\ell_4$ and at $\vec\ell = - \lambda
\vec\ell_5$ with $0 < \lambda < \infty$. As with $F_a$, as $\vec\ell$
approaches one of these singularities, $F_b$ approaches $-\infty$.

\begin{figure}[htbp]
\centerline{\DESepsf(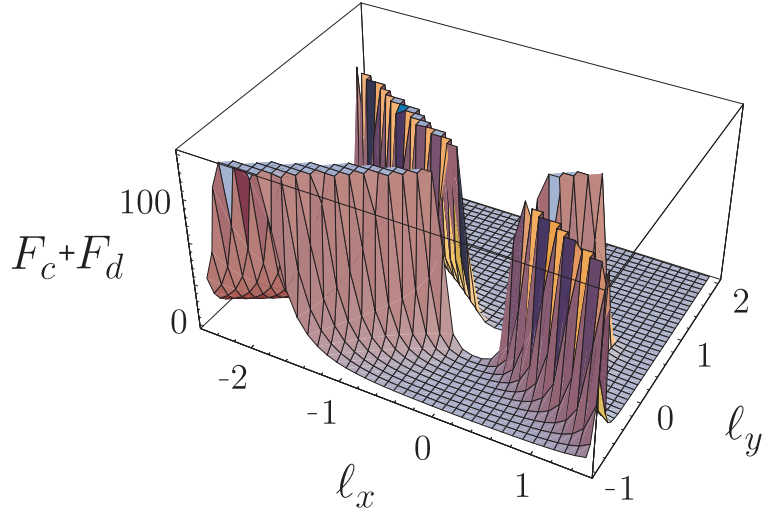 width 10 cm)}
\medskip
\caption{Contribution to $F$ from cuts $c$ and $d$. The same
rectangular domain as in Fig.~\ref{fig:Fa} has been removed from the
plot.  The function $F_c + F_d$ is singular along lines that extend
from $(1,0.5)$ to infinity in the direction $(-2,-1)$ and from
$(-2,1)$ to infinity in the direction $(2,-1)$. Only values $F_c + F_d
< 100$ are shown. }
\label{fig:Fcd}
\end{figure}

There are two cuts, $c$ and $d$, for which there are no virtual
subgraphs. In Fig.~\ref{fig:Fcd} I show the contribution $F_c +
F_d$ from these cuts. We see that $F_c + F_d$ approaches $+\infty$ at
just the singularities where $F_a$ and $F_b$ approach $-\infty$.

In the standard method for QCD calculations, we would perform the
integration over $\vec\ell$ partially numerically for the
contribution from cuts $c$ and $d$. Of course, we would have to do
something about the collinear and soft singularities, since otherwise
we would obtain an infinite result. For instance, if we were to use the
phase-space slicing method, we would slice away a small part of the
integration domain near the singularities and calculate its
contribution analytically in $3 - 2\epsilon$ spatial dimensions in the
limit that the region sliced away is small. Then we would be left with
a numerical integration of ${\cal G}_c + {\cal G}_d$ in the remaining
region (in exactly 3 spatial dimensions).  Evidently, the density of
points used in the present method would not do for this purpose; we
would need to expend more points on the region near the collinear
singularities.

\begin{figure}[htbp]
\centerline{\DESepsf(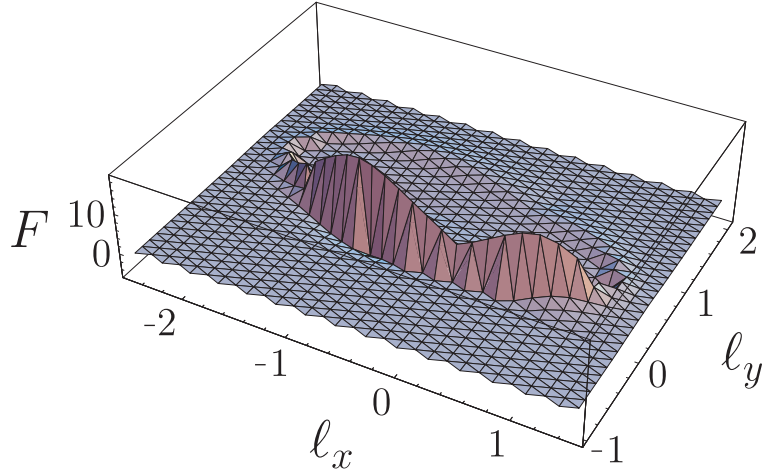 width 10 cm)}
\medskip
\caption{The net function $F$ with the contributions from cuts $a$,
$b$, $c$ and $d$ combined. There are no collinear singularities. What
remains after cancellation from the soft singularity is removed by the
density of points in the denominator of $F$. The remnants of the
scattering singularity are visible, but there is no actual singularity 
because of the contour deformation.}
\label{fig:Fabcd}
\end{figure}

We see that the standard method for performing the integrations, in
which some parts of the integrations are performed analytically and
some are performed numerically, is, of necessity, rather complicated.
In the numerical method, we simply combine  ${\cal G}_a$, ${\cal G}_b$,
${\cal G}_c$, and ${\cal G}_d$ and integrate numerically. The argument
in the preceding sections showed that the contributions from the
various cuts cancel as one approaches the collinear
singularities. This is illustrated in Fig.~\ref{fig:Fabcd}, where I
plot $F_a + F_b + F_c + F_d$ versus $\ell_{x}$ and $\ell_{y}$. We
see, first of all, that the singular behaviors at the collinear
singularities cancel, just as the calculation of
Sec.~\ref{sec:cancellation} showed. There is also a cancellation at
the soft singularity at $\vec\ell = 0$. There is still a singularity
in the integrand at $\vec\ell = 0$, but it is integrable and is
removed from $F$ by choosing a suitable density of points $\rho$.
Thus $F$ remains less than about 20 everywhere.

\begin{figure}[htbp]
\centerline{\DESepsf(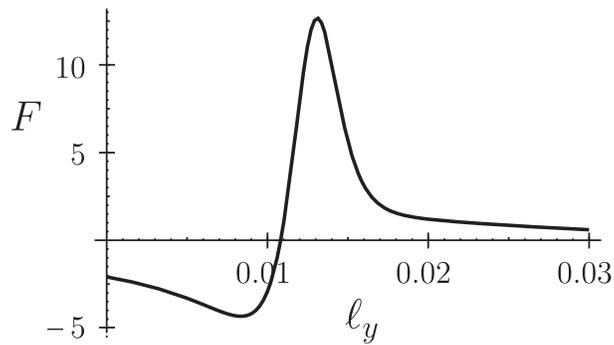 width 8 cm)}
\medskip
\caption{Structure of  $F$ along a slice through
Fig.~\ref{fig:Fabcd} at $\ell_{x} = -0.3$. Note that the ridge and
valley structure is very narrow.}
\label{fig:slice}
\end{figure}

We can see the remnants of the scattering singularity, which is
located on an ellipsoidal surface that intersects the plane
$\ell_{z} = 0$. If it were not for the contour deformation, $F$
would be singular on this surface, approaching $+\infty$ as one
approached the surface from one side and approaching $-\infty$ as
one approached the surface from the  other side. Since the deformed
contour avoids the singularity, the singular behavior is removed and we
are left with a ridge and valley near the ellipsoid. This structure is
illustrated in Fig.~\ref{fig:slice}, in which we see a slice through
the ridge and valley at $\ell_{x} = - 0.3$. Since the amount
of deformation vanishes as one approaches $\vec\ell = 0$, the width
of the ridge and valley structure becomes more and more narrow as
$|\vec\ell| \to 0$. Recall that the density $\rho$ of integration
points is designed to match this increasing narrowness as 
$|\vec\ell| \to 0$, so that the integration points are concentrated
where the structure is.

\section{Other issues}
\label{sec:other}

In the preceding sections, we have seen the most important features
of the method of numerical integration for one loop QCD calculations.
There are other important issues that are outside of the scope of this
paper. I mention these briefly here.

First, full QCD has a much more complicated structure than $\phi^3$
theory, which was the example for this paper. However, the
complications of full QCD are in the numerators of the expressions
representing Feynman diagrams, while the cancellations and the
analytic structure related to the contour deformation have to do with
the denominator structure. Thus one can simply generate the numerator
structure with computer algebra and carry it along.

\begin{figure}[htbp]
\centerline{\DESepsf(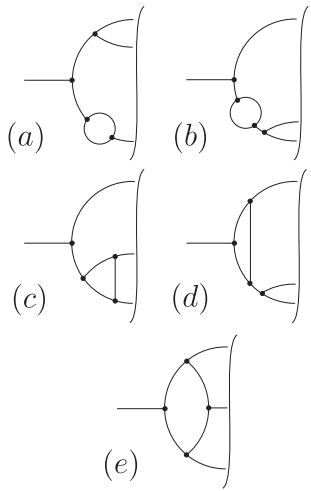 width 4 cm)}
\medskip
\caption{Ways to insert virtual subdiagrams in Feynman diagrams for
$e^+ e^- \to 3 {\ \it partons}$.}
\label{fig:virtualtypes}
\end{figure}

Second, the denominator structure in the example used in this paper is
not the only denominator structure that one needs to treat. In the QCD
calculation for three-jet-like quantities in electron-positron
annihilation at order $\alpha_s^2$, there are five possibilities for
how a virtual subgraph can occur inside an amplitude. The
possibilities are indicated in Fig.~\ref{fig:virtualtypes}. For each
possibility  there is an entering line representing the virtual
photon or $Z$ boson, which we take to have zero three momentum, and
there are three on-shell lines entering the final state. There are
graphs of two types, (a) and (b), containing two point virtual
subgraphs. There are graphs of two types, (c) and (d), containing
three point virtual subgraphs. There is one type, (e), of graph with a
four point virtual subgraph. In structure (c), a line with non-zero
three momentum enters the three point virtual subgraph and two
on-shell lines leave. This is the case that we analyzed in the example
of this paper. The structure of the graph led to the singularity
structure depicted in Fig.~\ref{fig:singularities}. Amplitudes of
types (d) and (e) have different singularity structures from that
studied here. Case (d) is simpler than the case we have studied, while
case (e) is somewhat more complicated.  However, the essential
features are those that we have already studied.

This leaves virtual self-energy subgraphs. In case (a), there is a self
energy subgraph on a propagator that enters the final state. This case
requires a treatment different from that discussed in the previous
sections. This is evident because there is a nominal $1/k^2$ where
$k^2 = 0$. The treatment required is to represent the virtual
self-energy via a dispersion relation \cite{beowulfprl}. In this
representation, the subgraph is expressed as an integral over the
three-momentum in the virtual loop with an integrand that is closely
related to the integrand for the corresponding cut self-energy graph.
The point-by-point cancellation between real and virtual graphs is
then manifest. It is convenient also to use the dispersive
representation for the much easier case (b).

Third, we have to do something about ultraviolet divergences in
virtual subgraphs. These are easily removed \cite{beowulfprl} by
subtracting an integrand that, in the region of large loop
momenta, matches the integrand of the divergent subdiagram. The
integrand of the subtraction term should depend on a mass parameter
$\mu$ that serves to make the subtraction term well behaved in the
infrared. Then, with the aid of a small analytical calculation for
each of the one loop divergent subdiagrams that occur in QCD, one can
arrange the definition so that this {\it ad hoc}  subtraction has
exactly the same effect as $\overline{\rm MS}$ subtraction with scale
parameter $\mu$.

Fourth, I use Feynman gauge for the gluon field, but then self-energy
corrections on a gluon propagator require special attention
\cite{beowulfprl}.  The one loop gluon self-energy subgraph,
$\pi^{\mu\nu}(k)$, contains a term proportional to $k^\mu k^\nu$ that 
contributes quadratic infrared divergences \cite{sterman}, while the
cancellation mechanism that we have studied in this paper takes care
of logarithmic divergences. This problem can be solved by
replacing  $\pi^{\mu\nu}$ by $P^\mu_\alpha\pi^{\alpha\beta}
P_\beta^\nu$, where $ P^\mu_\alpha = g^\mu_\alpha - { k^\mu \tilde
k_\alpha / \tilde k^2}$, with $\tilde k = (0,\vec k)$.  The terms
added to $\pi^{\mu\nu}$ are proportional to either $k^\mu$ or $k^\nu$
and thus vanish when one sums over different ways of inserting the
dressed gluon propagator into the remaining subgraph. Since
$P^\mu_\alpha k^\alpha = 0$, the problematic $k^\mu k^\nu$ term is
eliminated. Effectively, this is a change of gauge for  dressed gluon
propagators from Feynman gauge to Coulomb gauge.

Most of these issues are discussed briefly in \cite{beowulfprl}. A
quite detailed, but not very pedagogical, treatment can be found in
\cite{technotes}. Further analysis of these issues is left for a future
paper.

I have also not given a complete presentation of algorithms for
choosing integration points. As discussed in Sec.~\ref{subsec:nearsoft},
the crucial issue is to have the right singularities in the density of
points near a soft parton singularity of the Feynman diagram. This is
not the only issue that needs to be addressed in a complete algorithm.
Of course, the demonstration program \cite{beowulf} has a complete
algorithm. However, this algorithm is quite a hodge-podge of methods
and it seems that  a detailed exposition should be reserved for a
better and more systematic method, which remains to be developed.

\section{Results and conclusions}
\label{sec:conclusions}

In the preceding sections, we have seen some of the techniques needed
for the numerical integration method for QCD calculations.  Of
course, since not all of the techniques have been explained, the
explanation does not constitute a very convincing argument that such a
calculation is feasible. A truly exhaustive explanation would help,
but an actual computer program that demonstrates the techniques is
better. Results from such a program were presented in
\cite{beowulfprl}. Since then, I have found and corrected one bug that
resulted in errors a little bit bigger than 1\% and have made some
other improvements in the code. The resulting code and documentation
are available at \cite{beowulf}.

The program is a parton-level event generator. The user is to supply a
subroutine that calculates how an event with three or four partons in
the final state contributes to the observable to be calculated. The
program supplies events, each consisting of a set of parton momenta
$\{\vec k_1,\vec k_2,\vec k_3\}$ or $\{\vec k_1,\vec k_2,\vec k_3,\vec
k_4\}$, together with weights $w$ for the events. Then the user routine 
calculates ${\cal I}$ according to
\begin{equation}
{\cal I} \approx {1 \over N} \sum_{i = 1}^N w_i\ {\cal S}(k_i).
\end{equation}
The weights used are the real parts of complex weights; the imaginary
parts can be dropped since we always know in advance that ${\cal I}$
is real.  Thus the weights are both positive and negative. It would,
of course, be more convenient to have only positive weights, but one
can hardly have quantum interference without having negative numbers
along with positive numbers.

The first general purpose program for QCD calculation of
three-jet-like quantities in $e^+e^- \to {\it hadrons}$ at order
$\alpha_s^2$ was that of Kunszt and Nason \cite{KN}. This
program uses the numerical/analytical method of Ellis, Ross, and
Terrano. In Table~\ref{tab:thrust}, I compare the results of the
Kunszt-Nason program to those obtained with the numerical method for
the $\alpha_s^2$ contributions to moments of the thrust distribution,
\begin{equation}
{\cal I}_n = {1\over \sigma_0\, (\alpha_s/\pi)^2}
\int_0^1\! dt\ (1-t)^n
 {d\sigma^{[2]} \over dt}.
\label{tmoments}
\end{equation}
In Table~\ref{tab:ycut}, I compare the results of the two methods for
moments of the $y_{\rm cut}$ distribution for the three jet cross
section. To define these quantities, let $f_3(y_{\rm cut})$ be the
cross section to produce three jets according to the Durham algorithm
\cite{durham} with resolution parameter $y_{\rm cut}$. Let $g_3(y_{\rm
cut})$ be the negative of its derivative,
\begin{equation}
g_3(y_{\rm cut}) = -{f_3(y_{\rm cut})\over dy_{\rm cut}}.
\end{equation}
Then we calculate moments of the $\alpha_s^2$ contribution to this
quantity,
\begin{equation}
{\cal I}_n = {1\over \sigma_0\, (\alpha_s/\pi)^2}
\int_0^1\! d\,y_{\rm cut}\ (y_{\rm cut})^n
 g_3^{[2]}.
\label{ymoments}
\end{equation}
In each table, the results for the numerical method are shown with
their statistical and systematic errors. (The systematic error is
estimated by changing the cutoffs that remove small regions near the
singularities where roundoff errors start to become a problem.) The
corresponding results for the numerical/analytical method are shown
with the statistical errors as reported by the program. Inspection of
the tables shows that there is good agreement between the two methods.

\begin{table}
\caption{Comparison of results for moments of the thrust
distribution, Eq.~(\ref{tmoments}). The ``numerical'' results are from
the program \protect\cite{beowulf}. The first error is statistical, the
second systematic. The ``numerical/analytical'' results are from the
program of Kunszt and Nason \protect\cite{KN} and are given with their
reported statistical errors.}
\label{tab:thrust}
\medskip
\begin{tabular}{lll}
\multicolumn{1}{c}{n}  
&\multicolumn{1}{c}{numerical} 
&\multicolumn{1}{c}{numerical/analytical} \\
\tableline
1.5 & $\hspace{0.38 em}4.127 \pm  0.008 \pm 0.025$
    & $\hspace{0.38 em}4.132 \pm  0.003$          \\
2.0 & $\hspace{0.38 em}1.565 \pm  0.002 \pm 0.007$ 
    & $\hspace{0.38 em}1.565 \pm  0.001$          \\
2.5 &$(6.439 \pm  0.010 \pm 0.022)\times 10^{-1}$
    &$(6.440 \pm  0.003\hspace{0.5 em})\times 10^{-1}$\\
3.0 &$(2.822 \pm  0.005 \pm 0.009)\times 10^{-1}$
    &$(2.822 \pm  0.001\hspace{0.5 em})\times 10^{-1}$\\
3.5 &$(1.296 \pm  0.002 \pm  0.004)\times 10^{-1}$
    &$(1.296 \pm  0.0005)\times 10^{-1}$\\
4.0 &$(6.159 \pm  0.011 \pm 0.016)\times 10^{-2}$
    &$(6.161 \pm  0.002\hspace{0.5 em})\times 10^{-2}$\\
4.5 &$(3.009 \pm  0.006 \pm 0.007)\times 10^{-2}$
    &$(3.010 \pm  0.0006)\times 10^{-2}$\\
5.0 &$(1.501 \pm  0.003 \pm 0.003)\times 10^{-2}$
    &$(1.502 \pm  0.0002)\times 10^{-2}$\\
\end{tabular}
\end{table}

\begin{table}
\caption{Comparison of results for moments of the $y_{cut}$
distribution, Eq.~(\ref{ymoments}). The ``numerical'' results are from
the program \protect\cite{beowulf}. The first error is statistical, the
second systematic. The ``numerical/analytical'' results are from the
program of Kunszt and Nason \protect\cite{KN} and are given with their
reported statistical errors.}
\label{tab:ycut}
\medskip
\begin{tabular}{lrr}
\multicolumn{1}{c}{n}  
&\multicolumn{1}{c}{numerical} 
&\multicolumn{1}{c}{numerical/analytical} \\
\tableline
1.5 &$(8.442 \pm  0.034 \pm 0.059)\times 10^{-1}$
    &$(8.397 \pm  0.002\hspace{0.5 em})\times 10^{-1}$\\
2.0 &$(3.106 \pm  0.012 \pm 0.015)\times 10^{-1}$
    &$(3.090 \pm  0.0004)\times 10^{-1}$\\
2.5 &$(1.205 \pm 0.005 \pm 0.005)\times 10^{-1}$
    &$(1.200 \pm 0.0002)\times 10^{-1}$\\
3.0 &$(4.945 \pm 0.025 \pm 0.019)\times 10^{-2}$
    &$(4.927 \pm 0.001\hspace{0.5 em})\times 10^{-2}$\\
3.5 &$(2.122 \pm 0.012 \pm 0.008)\times 10^{-2}$
    &$(2.116 \pm 0.0007)\times 10^{-2}$\\
4.0 &$(9.430 \pm 0.064 \pm 0.032)\times 10^{-3}$
    &$(9.412 \pm 0.004\hspace{0.5 em})\times 10^{-3}$\\
4.5 &$(4.304 \pm 0.034 \pm 0.014)\times 10^{-3}$
    &$(4.301 \pm 0.002\hspace{0.5 em})\times 10^{-3}$\\
5.0 &$(2.008 \pm 0.018 \pm 0.006)\times 10^{-3}$
    &$(2.008 \pm 0.001\hspace{0.5 em})\times 10^{-3}$\\
\end{tabular}
\end{table}

We have explored some of the most important techniques necessary for a
QCD calculation for three-jet-like quantities in electron-positron
annihilation at order $\alpha_s^2$ using numerical integration
throughout the calculation. For the techniques covered, this
explanation expands on the brief presentation in \cite{beowulfprl}. We
have also seen that the method works. The older and very successful
numerical/analytical method for QCD calculations has its
complications. The numerical method has its own complications, but
they are different complications. Thus one may expect that the classes
of problems for which each of the methods is well adapted may be
different. There may be some classes of problems for which the natural
flexibility of the numerical method makes it the more useful method.
It remains for the future to explore the possibilities.

\acknowledgements

I thank M.~Seymour and Z.~Kunszt for providing advice and
results from their programs to help with debugging the program
described here. I thank P.~Nason for providing the Kunszt-Nason
program that I used in preparing the Tables \ref{tab:thrust} and
\ref{tab:ycut}. I thank M.~Kr\"amer for criticisms of the manuscript.
Finally, I am most grateful to the TH division of CERN for its
hospitality during a sabbatical year in which much of the writing
of this paper was accomplished.

\appendix
\section*{Contour deformation in many dimensions}

The calculational method described in this paper makes use of Cauchy's
theorem in a multi-dimensional complex space. Since this theorem is
not proved in most textbooks on complex analysis, I provide a proof
here, including the special case, needed for our application, in which
there is a singularity on the integration contour.

Let $f(z)$ be a function of $N$ complex variables $z^\mu = x^\mu + i
y^\mu$, with $\mu = 1,\dots,N$, where $x^\mu$ and $y^\mu$ are real
variables. Consider a family of integration contours ${\cal C}(t)$
labeled by a parameter $t$ with $0 \le t \le 1$ and specified by
\begin{equation}
z^\mu(x^1,\cdots,x^N;t) =
x^\mu + i\, y^\mu(x^1,\cdots,x^N;t),
\hskip 2cm \mu = 1,\dots,N.
\end{equation}
Let ${\cal I}(t)$ be the integral of $f$ over the contour ${\cal
C}(t)$,
\begin{equation}
{\cal I}(t) = \int_{{\cal C}(t)}\! d\,z\ f(z)
= \int\! d\,x\ \det\!\left(\partial z(x;t) \over \partial x\right)\
 f(z(x;t)).
\end{equation}
Suppose that $f(z)$ is analytic in a region that contains the
contours. Then we have {\em Cauchy's theorem}:
\begin{equation}
{\cal I}(1) = {\cal I}(0).
\end{equation}

To prove this theorem, we simply prove that $d\,{\cal I}(t)/d\,t = 0$.
Define
\begin{equation}
A^\mu_\nu = {\partial z^\mu \over \partial x^\nu}
=\delta^\mu_\nu + i\ {\partial y^\mu \over \partial x^\nu}.
\end{equation}
Let $B^\mu_\nu$ be the inverse matrix to $A^\mu_\nu$. Then
\begin{equation}
B^\mu_\nu \det A =
{1\over(N-1)!}
\epsilon^{\mu\mu_2\cdots\mu_N}
\epsilon_{\nu\nu_2\cdots\nu_N}\,
{\partial z^{\nu_2} \over \partial x^{\mu_2}}\cdots
{\partial z^{\nu_N} \over \partial x^{\mu_N}},
\end{equation}
where $\epsilon^{\mu_1\cdots\mu_N}$ is the completely antisymmetric
tensor with $N$ indices, normalized to $\epsilon^{1\,2\cdots N} = 1$,
and $\epsilon_{\mu_1\cdots\mu_N}$ is the same tensor. This has the
immediate consequence that
\begin{equation}
{\partial \over \partial x^\mu}\,
\left(B^\mu_\nu \det A \right)
= 0.
\label{rule1}
\end{equation}
Also,
\begin{equation}
\det A = 
{1\over N!}\
\epsilon^{\mu_1\cdots\mu_N}
\epsilon_{\nu_1\cdots\nu_N}\,
{\partial z^{\nu_1} \over \partial x^{\mu_1}}\cdots
{\partial z^{\nu_N} \over \partial x^{\mu_N}},
\end{equation}
so
\begin{equation}
{\partial \over \partial t}\det A
= {\partial  A^\nu_\mu\over \partial t}\ B^\mu_\nu \det A
= i\,{\partial  y^\nu\over \partial x^\mu\partial t}\ 
B^\mu_\nu \det A .
\label{rule2}
\end{equation}
We need one more result:
\begin{equation}
{\partial f \over \partial x^\mu}
= {\partial z^\nu \over \partial x^\mu}
{\partial f \over \partial z^\nu},
\end{equation}
so
\begin{equation}
{\partial f \over \partial z^\nu}
= B^\mu_\nu\ {\partial f \over \partial x^\mu}.
\label{rule3}
\end{equation}
Then, using the results (\ref{rule1}), (\ref{rule2}), and
(\ref{rule3}) and an integration by parts, we find
\begin{eqnarray}
{d \over d\,t}\,{\cal I}(t)
&=&
\int\!d\,x\ {d \over d\,t}\,\left[\det A\ f\right]
\nonumber\\
&=&
\int\!d\,x\ \det A \left\{
i\,{\partial  y^\nu\over \partial x^\mu\partial t}\ 
B^\mu_\nu \ f
+ {\partial f\over \partial z^\nu}\ 
i\,{\partial  y^\nu \over \partial t}
\right\}
\nonumber\\
&=&
\int\!d\,x\ \det A \left\{
i\,{\partial  y^\nu\over \partial x^\mu\partial t}\ 
B^\mu_\nu \ f
+i\, B^\mu_\nu\ {\partial f \over \partial x^\mu}\
{\partial  y^\nu \over\partial t}
\right\}
\nonumber\\
&=&
i\int\!d\,x\ B^\mu_\nu \det A\ 
{\partial \over \partial x^\mu}
\left\{
{\partial  y^\nu\over \partial t}\ f
\right\}
\nonumber\\
&=&
-i\int\!d\,x\ {\partial  y^\nu\over \partial t}\ f\
{\partial \over \partial x^\mu}
\left\{
B^\mu_\nu \det A 
\right\}
\nonumber\\
&=& 0.
\end{eqnarray}
This proves the theorem.

Consider now a more complicated problem. Suppose that we have an
integral of the form
\begin{equation}
{\cal I} = \int\! d\,x\ 
 [f(x) + g(x)].
\end{equation}
where $f$ and $g$ are both singular on a surface ${\cal P}$ in the
space of the real variables $x$. Suppose that the strength of the
singularities are such that the integral of either function would be
logarithmically divergent. Suppose further that there is a cancellation
in the sum such that the integral of $f+g$ is convergent. Let $d(x)$ be
the distance from any point $x$ to the surface ${\cal P}$. Let us cut
out a region of radius $R$ around ${\cal P}$ and write
\begin{equation}
{\cal I} = 
\lim_{R\to 0}\left[\int_{d>R}\! d\,x\ f(x)
 + \int_{d>R}\! d\,x\ g(x)\right].
\end{equation}
Now we wish to explore the consequences of deforming the integration
contour for the integral of $f$. Thus we investigate (with the same
notation as above)
\begin{equation}
{\cal I}(t) = 
\lim_{R\to 0}\left[\int_{d>R}\!
 d\,x\ \det\!\left(\partial z(x;t) \over \partial x\right)\
 f(z(x;t))
 + \int_{d>R}\! d\,x\ g(x)\right].
\end{equation}
Following the previous proof we find that there is a surface term in
the integration by parts
\begin{eqnarray}
{d \over d\,t}\,{\cal I}(t)
&=&
\lim_{R\to 0}\left[
i\int_{d<R}\!d\,x\
{\partial \over \partial x^\mu}
\left\{
 B^\mu_\nu \det A\ 
{\partial  y^\nu\over \partial t}\ f
\right\}\right]
\nonumber\\
&=& 
\lim_{R\to 0}\left[
i\int\!dS_\mu
\left\{
 B^\mu_\nu \det A\ 
{\partial  y^\nu\over \partial t}\ f
\right\}\right],
\end{eqnarray}
where the integration is over the surface $d = R$ and $dS_\mu$ is the
surface area differential normal to the surface.

We want to arrange the deformation specified by $y^\mu(x;t)$ so that 
$d{\cal I}(t)/dt = 0$. For this to happen, it is clear that $y$ will
have to approach 0 as $x$ approaches the surface ${\cal P}$. Then
$B^\mu_\nu \to \delta^\mu_\nu$ and  $\det A \to 1$ as $x$ approaches
${\cal P}$. Let the dimensionality of the singular surface ${\cal P}$
be $N-a$.  If the function $f$ was such that the original integral was
logarithmically divergent, then $f \propto R^{-a}$ for $R \to 0$. The
integration over the surface gives a factor $dS^\mu \propto R^{a-1}$
for $R \to 0$. Suppose that the deformation vanishes proportionally to
$R^b$. Then
\begin{equation}
{d \over d\,t}\,{\cal I}(t)
\propto
\lim_{R\to 0}\left[
R^{a-1}R^a R^b
\right].
\end{equation}
Then $d{\cal I}(t)/dt = 0$ if $b > 1$. The choice made in the main
text of the paper is $b = 2$.


\end{document}